\documentclass[a4paper,11pt]{article}
\usepackage{jheppub}
\pdfoutput=1

\usepackage{cancel}
\usepackage[normalem]{ulem}
\usepackage{float}
\usepackage{autobreak}
\usepackage{graphicx}
\usepackage{appendix}
\usepackage{amsfonts}
\usepackage{comment}
\usepackage{multirow}
\usepackage{slashed}
\usepackage{amsmath}
\usepackage{float}
\usepackage{autobreak}
\usepackage{graphicx}
\usepackage{appendix}
\usepackage{comment}

\usepackage{color,hyperref}
\usepackage[T1]{fontenc}
\usepackage{tocloft}
\usepackage[normalem]{ulem}
\usepackage{subcaption}
\usepackage{array}
\usepackage{enumitem}
\usepackage{extarrows}
\usepackage{mathtools}
\usepackage{slashed} 
\usepackage{color}
\usepackage{soul}
\sethlcolor{yellow} 

\newcommand{\nn}{\nonumber\\}

\newcommand{\Ca}{C_A}

\newcommand{\df}{{\rm d}}
\def\Dm1{{{\delta(1-z)}}}

\def\g0#1DY{{g_{0#1}^{DY}}}

\def\LogmW1{{{\ln (1-\omega)}}}

\newcommand{\overbar}[1]{\,\overline{\!{#1}}}
\newcommand{\Nbar}{\overbar{N}}

\newcommand{\as}{a_s}
\newcommand{\muf}{\mu_F}
\newcommand{\mur}{\mu_R}

\newcommand{\Lw}{L_\omega}
\newcommand{\wb}{\overbar{\omega}}

\newcommand{\eq}[1]{Eq.\ (\ref{#1})}

\newcommand{\tab}[1]{Table\ \ref{#1}}

\title{{Associated $ZH$ production in gluon fusion process at NLO+NLL}}
\author[a,b]{Pulak Banerjee,}
\emailAdd{pulak.banerjee@niser.ac.in}
\affiliation[a]{School of Physical Sciences, National Institute of Science Education and Research, Jatni, 752050, India}
\affiliation [b]{Homi Bhabha National Institute, Training School Complex, Anushakti Nagar, Mumbai, 400094, India}
\author[c]{Chinmoy Dey,}
\emailAdd{d.chinmoy@iitg.ac.in}
\affiliation[c]{Theoretical Physics Division, Physical Research Laboratory,\\ Navrangpura, Ahmedabad 380009, India}
\author[d]{Niraj Koirala}
\emailAdd{n.koirala@iitg.ac.in}
\affiliation[d]{Department of Physics, Indian Institute of Technology Guwahati,\\ Guwahati-781039, Assam, India}
\author[d]{M. C. Kumar}
\emailAdd{mckumar@iitg.ac.in}
\author[d]{and Vaibhav Pandey}
\emailAdd{vphiitg@iitg.ac.in}

\abstract{ We present precise results for invariant mass distributions and inclusive cross-sections, of associated $ZH$ production through gluon fusion in QCD. We include threshold logarithms at next-to-leading-logarithmic(NLL) accuracy and match the results to full next-to-leading-order(NLO) results with exact top-quark mass dependence in the virtual amplitudes.
At 13.6 TeV energy at LHC, the NLO+NLL cross-section increases the NLO counterpart by about $20\%$. For differential distributions, at 3000 GeV, the uncertainties arising out of the unphysical renormalization and factorization scales is about $20\%$ for NLO; whereas for the resummed results it is around $12\%$. We also combine these results to the Drell-Yan type results at N$^3$LL and present the most precise results in hadron collisions.
}

\begin{document}

\flushbottom
\keywords{Resummation, Higgs Physics}

\maketitle

\flushbottom

\tableofcontents



\section{Introduction}
\label{sec:intro}
Precise measurement of the Yukawa coupling is important to understand the origin of fermion masses. The decay of the Standard Model (SM) Higgs boson into bottom quarks is an important process which has been observed at ATLAS \cite{ATLAS:2018kot, ATLAS:2020fcp} and CMS \cite{CMS:2018nsn}.
	In the fermionic sector, the Higgs to $b\bar{b}$ is the largest being $58.2\%$ followed by decay to $\tau \bar{\tau}$ being $6.3\%$  and then to $c \bar{c}$ which is $2.9\%$~\cite{ATLAS:2024yzu}.
	Owing to its large branching fraction to $b \bar{b}$ the measurement of Yukawa coupling ($y_{b}$) is important to constrain new physics~\cite{ATLAS:2022vkf,CMS:2022dwd}.
	The associated production (also known as Higgs Strahlung) of Higgs with a vector boson ($W$ or $Z$) in which Higgs decay to $b \bar{b}$ and $Z$-boson decays to leptons is a clean signal as the leptonic decay of the $Z$-bosons suppresses the large QCD background. 
The quark-initiated Higgs Strahlung process starts at Leading order (LO), which is an electroweak process. Being a Drell-Yan like process at the LO, higher-order QCD corrections to the Higgs Strahlung process closely follow the DY process. Higher-order QCD corrections for Higgs Strahlung via DY process can be found in~\cite{Han:1991ia,Brein:2003wg,Brein:2012ne,Ferrera:2014lca,Campbell:2016jau,Ferrera:2017zex,Harlander:2018yio}. The unphysical scale uncertainty 
	in the inclusive cross-section at $13.6$ TeV reduces to $0.33\%$
	after including N$^3$LO corrections. 
	The NLO electroweak corrections to this process are known \cite{Ciccolini:2003jy, Denner:2011id, Denner:2014cla}; 
the corrections are around $-5\%$. 
Production of $ZH$ via bottom-quark annihilation in massless QCD has been studied in \cite{Ahmed:2019udm} and it contributes to sub-percent level accuracy. 
The next-to-next-to-leading-order (NNLO) corrections was investigated in \cite{Brein:2003wg}, the total production cross-section increases
about $3\%$ of the NLO cross-section
at the energies of LHC. 
The dependence on the unphysical renormalization and factorization scales also reduces from about 10\% at NLO to $2-3$ \% at NNLO\cite{Brein:2003wg}. 
The total production cross-sections at NNLO, for the Higgs Strahlung processes, are available in the public code \texttt{vh@nnlo}~\cite{Brein:2012ne,Harlander:2018yio}. 
Differential cross sections have been studied in \cite{Ferrera:2014lca, Campbell:2016jau, Ferrera:2017zex}.
For an on-shell Higgs production in  association with a leptonically decaying gauge boson and a jet, see \cite{Gauld:2021ule}.
Study of $ZH$ production through quark anti-quark annihilation, with the Higgs boson being emitted from a massive top quark loop can be found in \cite{Brein:2011vx}. At the energies of LHC, the corrections are of the order of 1-2\%.
 
The resummation of large soft-virtual (SV) logarithms has been studied extensively \cite{Sterman:1986aj, Catani:1989ne, Catani:1990rp,Kidonakis:1997gm,Kidonakis:2003tx,Moch:2005ba,Laenen:2005uz,Kidonakis:2005kz,Ravindran:2005vv,Ravindran:2006cg,Idilbi:2006dg,Becher:2006mr,Ahmed:2020nci}. 
This framework has been used for several color-singlet production processes \cite{Catani:2003zt,Moch:2005ky,deFlorian:2007sr,Kidonakis:2007ww,Catani:2014uta,Bonvini:2014joa,
Ahmed:2015sna,Schmidt:2015cea,Ahmed:2016otz,Bonvini:2016frm,Kidonakis:2017dmh,AH:2019phz,Das:2019btv,Das:2019bxi,Das:2020gie,Das:2020pzo,AH:2020cok,AH:2022elh,Banerjee:2024xdh,Das:2025qym,Banerjee:2025tbo,Goyal:2025bzf,Goyal:2025jgt}, 
where it improves predictions for inclusive cross sections and invariant-mass distributions.
The results for $VH$ production process in the $q\bar{q}$ channel have been studied at NNLO+N$^3$LL in ~\cite{Dawson:2012gs}.
The threshold corrections beyond NNLO, arising out of soft and collinear modes of matrix amplitudes (soft virtual corrections (SV)), have been investigated in \cite{Kumar:2014uwa}. 
The change in total production cross-section is around $-0.06\%$ at the accuracy of N$^3$LO$_{\rm SV}$ for $13$ TeV LHC . 
The fully inclusive cross-section for the Higgs Strahlung process at N$^3$LO has been recently reported in a public code \texttt{n3loxs}~\cite{Baglio:2022wzu}; corrections at N$^3$LO are found to be 
$-0.9\%$ (at $13.6$ TeV), and the scale uncertainties are larger compared to NNLO~\cite{Baglio:2022wzu}.
Higher order effects in invariant mass distribution, beyond N$^3$LO, have been investigated via threshold resummation at N$^3$LO + N$^3$LL accuracy in \cite{Das:2022zie}. 
For the invariant mass distribution, the resummation effects reduce 
the seven-point scale uncertainties to about $0.1\%$ for $Q > 1500$ GeV. 
At the level of the total production cross-section, the unphysical scale uncertainties are $0.25\%$ larger at N$^3$LO + N$^3$LL accuracy than at the fixed-order counterpart~\cite{Das:2022zie}.
The fixed-order QCD along with parton shower effects have been investigated in \cite{Granata:2017iod,Luisoni:2013cuh,Goncalves:2015mfa,Hespel:2015zea,Astill:2018ivh,Bizon:2019tfo,Alioli:2019qzz}. \\
At NNLO of $pp \rightarrow ZH$, the loop-mediated process $g g \rightarrow ZH$ starts to contribute. Owing to the large gluon flux at the LHC, the process $g g \rightarrow ZH$ provides about 6\% of the total cross-section at 13 TeV~\cite{Chen:2022rua}. As a probe for new physics scenarios, the 
$g g \rightarrow ZH$ process plays an important role, see \cite{Englert:2016hvy,Yan:2021veo}. In addition, the $g g \rightarrow ZH$ process can be used to probe the $Zb\bar{b}$ coupling, which may help to understand the discrepancy between the bottom quark forward-backward asymmetry at LEP and the SM predictions \cite{Yan:2021veo}. By comparing the $ZH$ production mode to the $WH$ production (which has only a DY-like contribution) and the corresponding invariant mass distribution of $VH$, the authors in \cite{Harlander:2018yns} proposed a data-driven method to extract the $gg \rightarrow Z H$ contributions.  Being a non Drell-Yan like process with loop induced at LO, $g g \rightarrow ZH$ differs in terms of differential distributions from the Drell-Yan $q\bar{q}$ counterpart \cite{Ferrera:2014lca,Hespel:2015zea}. 
The LO of the gluon-initiated sub-process, with exact quark masses in the loop, was studied in \cite{Dicus:1988yh,Kniehl:1990iva,Kniehl:1990zu,Kniehl:2011aa}. The unphysical scale uncertainties are about 25\%; this induces about 3\% uncertainty on the full $ZH$ contribution~\cite{LHCHiggsCrossSectionWorkingGroup:2016ypw}. Higher-order corrections to the associated Higgs production in the gluon-initiated channel are important for the Higgs boson search. Study of boosted Higgs boson in gluon-initiated associated production forms an important search channel for Higgs at LHC \cite{Englert:2013vua}.
 In the boosted Higgs regime, large correction factors beyond LO, for this sub-process have been reported in \cite{Altenkamp:2012sx, Harlander:2013mla,Bizon:2021rww}. 
 Experimental measurements of Higgs Strahlung suffer from systematic uncertainties \cite{ATLAS:2019yhn}. The uncertainties in the predictions were about 2-3 times larger for $ZH$ than for $WH$, because of the $g g \rightarrow ZH$ sub-process. The large scale uncertainties play a significant role in the experimental analysis \cite{ATLAS:2018kot, CMS:2018nsn, ATLAS:2020jwz}. The LO scale uncertainties often do not capture the size of higher-order perturbative corrections.
It is thus important to reduce the theoretical uncertainties at LO by studying the NLO QCD corrections. 
 Calculating the exact NLO corrections with full massive-quark contributions is challenging due to the presence of multiple scales in the virtual amplitudes. Using the fact that $gg \rightarrow H$ and $gg \rightarrow ZH$ processes are similar in the QCD structure, the authors in \cite{Altenkamp:2012sx} studied the NLO cross-section (NLO-EFT) in the limit of infinite top-quark and vanishing bottom-quark masses. The scale uncertainties at NLO are found to be still large but a little less than those of LO.

Soft gluon effects at NLO + NLL for $ZH$ production in the EFT approach has also been considered in \cite{Harlander:2014wda}, where the authors perform resummation in two different threshold definitions at the level of total production only. One is the invariant-mass-threshold approach ($Q$ approach), where the threshold logarithms arising from soft-gluon effects are resummed with respect to the invariant-mass scale. The second approach ($M$ approach) consists of defining the threshold with respect to the minimum mass of the $ZH$ system. The $Q$ approach has been found to be better behaved than the $M$ approach in reducing scale uncertainty. 
 Later on, in the experimental analysis for $ZH$ production, \cite{ATLAS:2020jwz,ATLAS:2020fcp}, the NLO + NLL EFT cross-sections for the $gg \rightarrow ZH$ channel were taken into account. In \cite{Das:2025wbj}, the authors improved upon the NLO EFT results for  $gg \rightarrow ZH$ by rescaling NLO EFT K-factors with the exact LO result.
 More recently, next-to-soft (NSV) threshold contributions has been investigated
\cite{Kidonakis:1996aq,Moch:2009hr,Soar:2009yh,Bonocore:2015esa,DelDuca:2017twk,Beneke:2018gvs,Bahjat-Abbas:2019fqa,Beneke:2019oqx,Beneke:2019mua,Moult:2019mog,Liu:2020tzd,vanBeekveld:2019cks,Das:2020adl,AH:2020iki,AH:2021vdc,vanBeekveld:2021hhv,AH:2022lpp,Beneke:2022obx,Liu:2022ajh,Sterman:2022lki,Pal:2023vec,Banerjee:2024cpr,Bhattacharya:2025rqk,Das:2025wbj}.
  In the low invariant mass regions, the soft virtual cross-sections (next-to-soft virtual cross-sections) contribute an additional 19\% (35.3\%) at NLL accuracy ($\overline{\rm NLL}$).
  The conventional seven-point scale uncertainties are reduced by about 5\% (1.4\%) after the SV (NSV) resummation.

 It is known that, in general, loop-induced processes are sensitive to new physics. A full calculation at NLO, with exact top-quark masses in the loop, is essential. Using an asymptotic expansion in the top quark mass, approximate NLO results were presented in \cite{Hasselhuhn:2016rqt}.
 A few of the current authors made an attempt to obtain the invariant mass distribution for the $ZH$ production using the Born improved theory approach where the exact LO results were rescaled by NLO EFT K-factors. Subsequently the resummation results for the invariant mass distrinution as well as the total production cross section have also been presented in the born improved theory.
 In \cite{Davies:2020drs}, the virtual amplitudes were computed using a high-energy expansion, namely in $m_t^2 << s,t$, along with Padé approximation. The results agree with an exact numerical study \cite{Chen:2020gae}.  The inclusive and fully differential cross-sections have been studied using a power series expansion in $m_H^2$ and $m_Z^2$ in \cite{Wang:2021rxu} and also by using a fully numerical approach in \cite{Chen:2022rua}. Using a transverse-momentum expansion and a high-energy expansion, the $gg \rightarrow ZH$ virtual corrections were presented in \cite{Alasfar:2021ppe}; by including the full real-emission contributions, the full NLO QCD results were presented in \cite{Degrassi:2022mro,CampilloAveleira:2025rbh}. In \cite{Davies:2025out}, expansions around the forward and high-energy limits were performed. The unphysical scale variations at NLO, with exact top quark masses in all amplitudes, is still around 20\% \cite{Degrassi:2022mro}. This points to the unaccounted higher order contributions in the prediction for $gg \rightarrow ZH$ process. Recently, the planar master integrals for two loop electroweak contributions  has been reported in \cite{Li:2026emp}.  \\

Going beyond NLO accuracy for  $gg \rightarrow ZH$ process is difficult. The first steps have been taken in \cite{Davies:2025otz}; where $1/m_t$ expansion has been performed to achieve the results. Although NLO+NLL results are available in the born improved EFT result~\cite{Das:2025wbj}, yet investigation of the effect of exact top quark mass in the differential distributions, is essential. In this article, we take the first steps to achieve these results. We shall show the differences in the result between NLO+NLL EFT approach and that of ours.  To achieve this, we present total cross-section and invariant mass distribution for $gg \rightarrow ZH$ process, at NLO + NLL accuracy, with exact top quark mass dependence in the virtual amplitudes taken from \cite{Davies:2020drs,Davies:2025otz}. By including the results from $q\bar{q}$ channel at ${\mathcal O(\alpha_s^3)}$ accuracy, we present the most accurate results in QCD, for $pp \rightarrow ZH$ process. 
We have implemented Catani-Seymour subtraction at NLO in a private code and validate the existing distributions and cross-sections implemented in the recent public code \texttt{ggxy}~\cite{Davies:2026uxl,Davies:2025qjr}.\\

Our article is structured as follows, in section ~\ref{sec:theory} we give the formalism that is followed to perform the resummation. In section ~\ref{sec:results} we present the phenomeological predictions for the invariant-mass distribution as well as the total production cross-section for the gluon initiaed $ZH$ productiopn at the LHC energies, and finally we conculde in sec.\ref{sec:conclusion}.



\section{Theoretical Framework} \label{sec:theory}
In proton-proton collisions, the hadronic cross-section for $Z H$ production can be expressed using QCD factorization as,
\begin{align}\label{eq:had-xsect}
 \frac{\df \sigma}{\df Q}  
=  
\sum_{a,b}\int_0^1 \df x_1\int_0^1 \df x_2 \,\,f_{a}(x_1,\mu_F^2)\,
f_{b}(x_2,\mu_F^2) 
\int_0^1 \df z~ \delta \left(\tau-zx_1 x_2\right)
\frac{\df \widehat\sigma_{ab}(z,\muf^2)}{\df Q} \, ,

\end{align}
%
where $f_{a,b}$ are the parton distribution functions (PDFs) for parton $a,b$ in the incoming hadrons
and $\widehat\sigma_{ab}$ is the partonic coefficient function. 
The variables $ z= Q^2/\widehat{s}$ and $\tau=Q^2/S$ are the threshold variables,
 where $S$ is collider center-of-mass energy and $\widehat{s}$ is the partonic center-of-mass energy. 
Here, $Q$ is the invariant mass of the final state $Z H$ and the factoriation scale is represented as $\muf$.
The partonic coefficient function can be decomposed in terms of the soft-virtual(SV) and regular terms (suppressing all scale dependencies) as,
%
\begin{align}\label{eq:PARTONIC-DECOMPOSE}
\frac{\df \widehat\sigma_{ab}(z)}{\df Q}
& =
\widehat{\sigma}^{(0)}_{ab}(Q^2) \Big( 
\delta_{b {a}}\Delta_{ab}^{\rm SV}\left(z\right) 
+ \Delta_{ab}^{\rm REG}\left(z\right)
\Big) \,.
\end{align}
%
The {\rm SV} part ($\Delta_{ab}^{\rm SV}$) captures the leading singular contributions in the $z \to 1$ limit which typically arise from soft gluon radiations and virtual corrections. The SV terms gets contributions from the diagonal channels ($gg$ and $q\bar{q}$) only.
The $\Delta_{ab}^{\rm REG}$ term contains subleading or regular contributions in the variable $z$ which gets contributions from off-diagonals ($qg$) channels also. 
The $\Delta_{\rm SV}^{ab}$  and $\Delta_{\rm REG}^{ab}$ can be organized as perturbative expansion in the strong coupling $\alpha_s(\mur^2) \equiv 4 \pi a_s(\mur^2)$ at the renormalisation scale ${\mur}$ as, 
%
\begin{align}\label{eq:SVREG-Expansion}
\Delta^{\rm SV}_{ab}(z) 
&= \sum_{n=0}^{\infty}\as^n(\mur^2) ~\delta_{ab}
\left( 
\Delta^{(n)}_{\delta} ~\delta(1-z) + 
\sum_{k=0}^{2n-1}  \Delta^{(n)}_{{\cal D}_k} ~{\cal D}_k(1-z) 
\right)\,, ~ 
\text{with } ab \in \{gg, q\bar{q} \} \,,
\nonumber
\\
\Delta^{\rm REG}_{ab}(z) 
&= 
\sum_{n=0}^{\infty}\as^n(\mur^2)
\Delta_{ab}^{(n)}(z)\,,

\end{align}
where $\delta(1-z)$ is the Dirac delta distribution and ${\cal D}_k(1-z) \equiv \left[\ln^k(1-z)/1-z\right]_+$ are 
the plus-distributions. 

The singular SV part of the partonic coefficient has a universal structure it gets contributions from the 
underlying hard form factor,
mass factorization kernels 
\cite{Moch:2004pa,Vogt:2004mw} 
and soft function 
\cite{Ravindran:2005vv,
Ravindran:2006cg,
Sudakov:1954sw,
Mueller:1979ih,
Collins:1980ih,
Sen:1981sd}. 
The underlying hard formfactor and soft function are infrared divergent, when combined alongside with the mass-factorization kernels the SV cross section becomes finite.
These large distributions appearing in $\Delta^{\rm SV}$ can be resummed to all orders in the threshold limit $z \to 1$. 
Resummation is often performed in the Mellin $N$-space where plus-distributions become simple logarithms in Mellin variable ($N$). 
The threshold limit $z \to 1$ translates into $N \to \infty$ limit. 
The partonic SV coefficients in the Mellin space can be organized as follows,
\begin{align}\label{eq:resum-partonic}
	\frac{\df \widehat{\sigma}_{N,ab}^{\rm N^{\it n}LL}}{\df Q}
	= \int_0^1 \df z ~ z^{N-1} \left(  \Delta^{\rm SV}_{ab}(z) \right)
	\equiv g_{0}(Q^2) \exp \left( G_N \right)\,.
    
\end{align}
The factor $g_{0}$ is independent of the Mellin variable and contains process-dependent information.
The leading threshold enhanced large logarithms in Mellin space are resummed through the exponent $G_{N}$.
The resummed accuracy is determined through the successive terms from both the exponent $G_N$ which takes the form,
\begin{align}\label{eq:GN}
G_N &= 
	\ln (\Nbar) ~g_1(\omega)
        + \sum_{n=1}^{\infty} \as^{n-1}(\mur^2)~ g_{n+1}(\omega) \,,
\end{align}
where $\Nbar = N \exp (\gamma_E)$, with $\gamma_E$ denoting the Euler--Mascheroni constant, and \\ $\omega = 2 \beta_0 \as(\mur^2) \ln \Nbar$.
The leading term in the expansion of $G_N$ gives the leading-logarithmic (LL) contribution, while the addition of higher-order terms systematically improves the logarithmic accuracy up to NLL.
The coefficients $g_n$ are universal in the sense that they depend only on whether the initiating partons are quarks or gluons.
In the same way, achieving higher logarithmic accuracy requires retaining further terms in the expansion of $G_N$.
Starting at NLL, one must also include the $N$-independent coefficients $g_0$, whose perturbative expansion can be written as
\begin{align}
	g_0(Q^2) = 1+\sum_{n=1}^{\infty} \as^n(\mur^2) ~g_{_{0n}}(Q^2) \,.
    
\end{align}
%
The explicit expression for the SV resummation exponent can be found, for example, in \cite{Catani:2003zt,Moch:2005ba}.
To achieve NLL accuracy, one needs the following coefficients,
\begin{align}
    g_1(\omega) &= 
    \frac{1}{\beta_0} \bigg\{ 
                             C_A \left( 8 \left( 1+  \frac{\wb}{\omega}\Lw \right) \right)
                    \bigg\}
    \,,
    \nn
    g_2(\omega) &= \frac{1}{\beta_0^3} \bigg\{ 
            C_A \beta_1 \left( 4 \omega + 4 \Lw + 2 \Lw^2\right)
            + C_A n_f \beta_0 \left(\frac{40}{9} \left( \omega + \Lw\right) \right)
            \nn
            &\hspace{2.5em}+ C_A^2 \beta_0 \left( 
                                    \left(-\frac{268}{9} + 8 \zeta_2 \right) \left(\omega + \Lw\right) 
                            \right) 
            + C_A \beta_0^2 \left( 4 \Lw \ln  \left(\frac{Q^2}{\mur^2}\right) + 4 \omega \ln\left( \frac{\muf^2}{\mur^2}\right) 
 \right)
    \bigg\},
\end{align}
where $\wb = 1-\omega$ and $\Lw = \ln(\wb)$.
Up to $\text{NLL}$ accuracy, one also needs the process-dependent coefficient $g_{_{01}}$ which for the gluon subprocess is found to be~\cite{Ahmed:2020nci},
\begin{align}\label{eq:g01}
{g}_{_{01}}(Q^2) &= 2 \mathcal{M}_{(0,1)} 
        - 2 \beta_{0} \ln\left(\frac{\muf^2}{\mur^2}\right) 
        + 2 \Ca \ln\left(\frac{Q^2}{\mur^2}\right) 
        + 2 \Ca \zeta_{2}
\end{align}
Note that both $g_0$ and $G_{N}$ depend on the resummation scheme. This is connected to the freedom in exponentiating some constant terms, such as $\gamma_E$, that arise from the Mellin transformation together with the large-$N$ contributions; see, for example, \cite{AH:2020cok} for a detailed discussion.
For LHC applications, it has been found ~\cite{AH:2020cok} that the so-called $\overbar{N}$-scheme leads to a faster perturbative convergence of the resummed series.
In this scheme, the constant $g_0$ does not depend on $\gamma_E$.
The resummed expression in \eq{eq:resum-partonic} must finally be matched to the available fixed-order result.
This matching includes the hard regular contribution and avoids double counting of the SV logarithms.
It is usually carried out using the \textit{minimal prescription} \cite{Catani:1996yz}. In this method, we choose the contour as $N = c + x \exp{i\phi}$, where $x$ is a real variable. The parameter $c$ is chosen so that all singularities, except the Landau pole, lie to the left of the contour. For the numerical analysis, we use $c = 1.9$ and $\phi = 3\pi/4$. In this way, we obtain
\begin{align}\label{eq:MATCHING}
	\frac{\df \sigma^{\rm N^{\it n}LO}_{ab}}{\df Q}
	=&
	\frac{\df {\sigma}^{\rm N^{\it n}LO}_{ab} }{\df Q}
	+
        \sum_{ab \in \{gg, q\bar{q}\} }
        \widehat{\sigma}^{(0)}_{ab}(Q^2)
	\int_{c-i\infty}^{c+i\infty}
	\frac{\df N}{2\pi i}
	\tau^{-N}
	f_{a,N}(\muf)
	f_{b,N}(\muf)
	\nn
	&\times 
	\Bigg( 
		\frac{\df \widehat{\sigma}_{N,ab}^{\rm N^{\it n}LL}}{\df Q} 
		- 	
		\frac{\df \widehat{\sigma}_{N,ab}^{\rm N^{\it n}LL}}{\df Q} \Bigg|_{\rm tr}
	\Bigg) \,.
    
\end{align}
The quantities $f_{a,N}$ are the Mellin moments of the PDFs, defined in the same way as the partonic coefficient in \eq{eq:resum-partonic}.
They can be evolved, for example, with the public code QCD-PEGASUS \cite{Vogt:2004ns}.
For practical calculations, one may also approximate them by using the PDFs directly in $z$ space, following \cite{Catani:2003zt,Catani:1989ne}.
The subscript `$\text{tr}$' in the last term inside the brackets in \eq{eq:MATCHING} means that the resummed partonic coefficient in \eq{eq:resum-partonic} is expanded only up to the same fixed order.
This is done to avoid double counting terms that are already included in the fixed-order contribution through $\sigma_{ab}^{\rm N^{\it n}LO}$.
For SV resummation, this truncated piece contains all singular logarithms that are already present at fixed order.
In the next section, we discuss the impact of threshold resummation on the gluon subprocess at the LHC.


\section{Numerical results }\label{sec:results}

Here we present numerical results for the $ZH$ associated production at the LHC. We follow the same setup as the one in Ref.~\cite{Das:2025wbj}
. The default center-of-mass energy of the incoming protons  is set to be $13.6$ TeV. 
The strong coupling constant is supplied by the \texttt{LHAPDF} \cite{Buckley:2014ana} routine. The fine structure constant is taken as
$\alpha \simeq 1/127.93$. The masses of the weak gauge bosons are set to be $m_Z = 91.1880$ GeV and $m_W = 80.3692$ GeV 
\cite{ParticleDataGroup:2024cfk}. 
The Weinberg angle 
is related to the masses of the weak gauge bosons as $\text{sin}^2\theta_\text{w} = (1 - m_W^2/m_Z^2)$, which corresponds to
the weak coupling $G_F \simeq 1.204399\times 10^{-5} \text{ GeV}^{-2}$. 
We use $m_{H} = 125.2$ GeV for the Higgs boson mass.
Our numerical analysis employs the {\tt PDF4LHC21\_40} \cite{PDF4LHCWorkingGroup:2022cjn} parton 
distribution functions (PDFs) throughout, as provided by {\tt LHAPDF} \cite{Buckley:2014ana}
with the central pdf set being the standard choice.
The gluon fusion process starts contributing at $\mathcal{O}(\alpha_s^2)$ accuracy; which is the same order as the NNLO correction to the Drell-Yan process. 
Therefore, to maintain the consistency in the perturbative expansion, 
we have used the NNLO PDF for the computation of LO and higher order corrections to the gluon fusion process. 
%
For the box diagram contributions, we consider only the top quark contribution, and no massive bottom quarks are present in the loop.
The top quark pole mass is set to $m_t = 172.57$ GeV while bottom quark is treated as massless.
The unphysical renormalization ($\mur$) and factorization scales ($\muf$) are set to the invariant mass ($Q$) of the $ZH$ system 

For the scale uncertainty estimation we choose Q as the central scale and we vary both these unphysical scales 
in the range $[Q/2, 2Q]$ keeping the constraint 
$|\,\text{ln}(\mu_R/\mu_F)\,| < \text{ln}\,4$ (known as the 7-point scale variations) 
 and take the maximum absolute deviation of 
the cross-section from that obtained with the central scale choice. To estimate the impact of the higher order corrections from FO and
resummation, we follow the same notation as in \cite{Das:2025wbj}, given below for the convenience of the reader. The results are present in terms of the ratios
\begin{align}
        { K}_{nm}
        =
        \frac{\sigma^{\text{N}^{\it n}\text{LO}}}{\sigma_c^{\text{N}^{\it m}\text{LO}}}
         \text{ and }
        {R}_{nm}
        =
        \frac{\sigma^{\text{N}^{\it n}\text{LO} + \text{N}^{\it n}\text{LL}}}{\sigma_c^{\text{N}^{\it m}\text{LO}}} 
        \label{eq:ratio}
        \end{align}
Here subscript `$c$' given in above expressions indicates the quantity computed with the central scale choice. 

The NLO cross-sections are evaluted by implementing the Catani-Seymour dipole subtraction method~\cite{Catani:1996vz} using our in-house numerical routines.
The loop induced born ($gg \to ZH$) and real-emission amplitudes ($gg \to ZHg$) along with the necessary ingredients such as color and spin-correlated born amplitudes are taken from {\tt Recola2 }\cite{denner2018recola2}.
Other real emission amplitudes for subprocesses such as $qg \to ZHq$, $\bar{q}g \to ZH\bar{q}$ and $q\bar{q}\to ZHg$ are evaluated using { \tt MadGraph5\_aMC@NLO} package\cite{Alwall_2011}, where we have used a filter 
to remove DY like contributions.
The two-loop virtual amplitudes are taken from {\tt ggxy} package \cite{Davies:2025qjr,Davies:2025otz}, and the phase space integrations were subsequently performed using our in-house code. 

To validate our implementation of Catani-Seymour subtraction~\cite{Catani:1996vz}, we compare the total production cross-section in \tab{tab:code_comparison_tot} as well as the invariant mass distribution in \tab{tab:code_comparison_inv} 
to the corresponding ones obtained from
{\tt ggxy} package.
We compare over a wide range of center-of-mass energies and invariant mass values and note that the results agree to per-mille level accuracy. 
At NLO, there are a class of diagrams for $qg$ initiated process, where the Higgs boson is being emitted from a closed fermion loop, and the final state $Z$-boson is emitted from an external fermion line. 
These constitue the $Z$-radiated diagrams ($4$ in total for one flavor), which in high $Q$ region gives significant contribution to the K-factor, as observed in Ref.~\cite{Degrassi:2022mro}. 
We shall also include these contributions in our prediction for the soft-gluon resummation.
Further we also give predictions using the filter that removes these $Z$-radiated diagrams.

\begin{figure}[!htbp]
\centering
\includegraphics[width=0.6\textwidth]{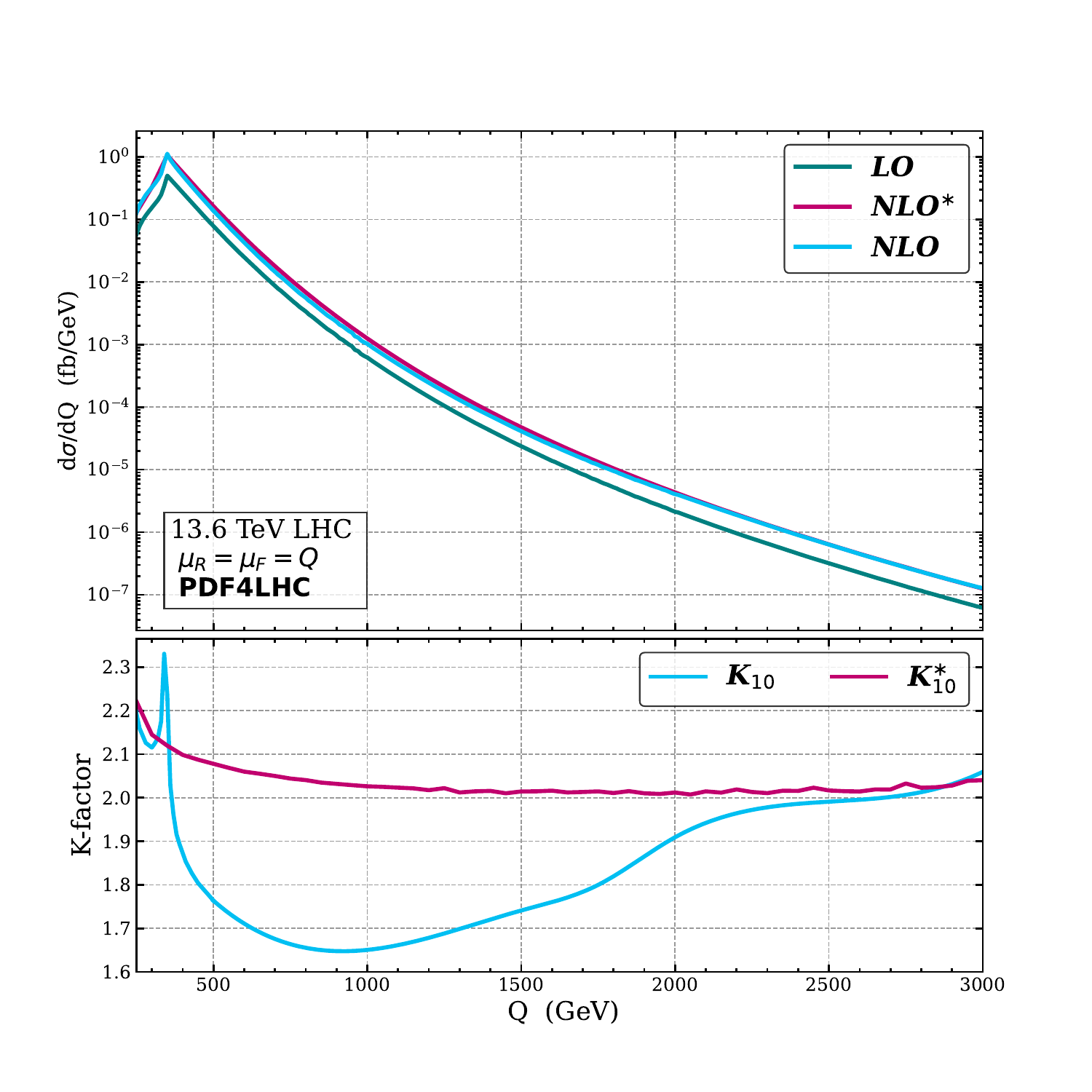}
\caption{
\small{
        Invariant mass distribution for the gluon fusion channel at NLO with the exact two-loop virtual amplitude (NLO) and the born-improved NLO correction (NLO$^{\ast}$). The respective K-factors are shown in the lower panel of the figure. 
}
}\label{fig:InvMass_wBIT}
\end{figure}

\begin{figure}[!htbp]
\centering
\includegraphics[width=0.6\textwidth]{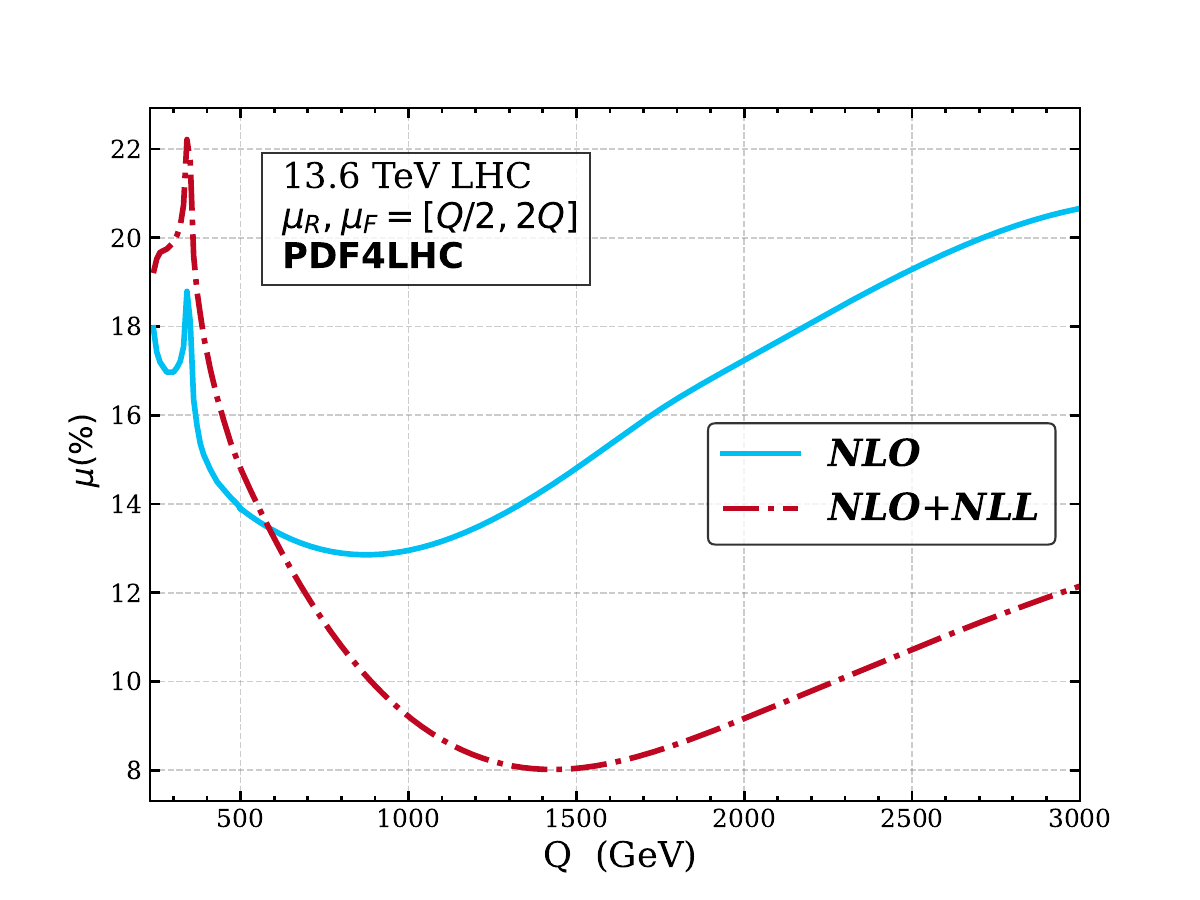}
\caption{
\small{
        The maximum percentage uncertainty from the 7-point scale variation for the invariant mass distribution (without the $Z$-radiated diagrams) at NLO and NLO$+$NLL accuracy.
        }
}\label{fig:7pt_unc}
\end{figure}

\begin{table}[!htbp]
	\begin{center}{
	\setlength{\extrarowheight}{5pt}
	\scalebox{0.92}{\begin{tabular}{|c|c|c|c|c|}
		\hline
		$\sqrt{s}$ (TeV) & In-house $\sigma_{LO}$ (fb) & {\tt ggxy} $\sigma_{LO}$  (fb) & In-house $\sigma_{NLO}$ (fb) & {\tt ggxy} $\sigma_{NLO}$ (fb) \\
		\hline
		13 & $54.53 \, ^{+24.96\%}_{-18.94\%}$ & $54.52 \, ^{+24.96\%}_{-18.94\%}$ & $108.47 \, ^{+16.25\%}_{-13.70\%}$  & $108.50 \, ^{+16.24\%}_{-13.75\%}$ \\
		\hline
		13.6 & $60.35 \, ^{+24.56\%}_{-18.72\%}$ & $60.35 \, ^{+24.57\%}_{-18.72\%}$ & $119.49 \, ^{+16.00\%}_{-13.60\%}$ & $119.82 \, ^{+16.03\%}_{-13.58\%}$ \\
		\hline
		14.0 & $64.38 \, ^{+24.32\%}_{-18.58\%}$ & $64.37 \, ^{+24.32\%}_{-18.58\%}$ & $127.28 \, ^{+15.89\%}_{-13.49\%}$ & $127.64 \, ^{+15.91\%}_{-13.49\%}$ \\
		\hline
	\end{tabular}
	}}
	\caption{
		\small{Comparison of total $ZH$ production cross-sections at LO and NLO (in fb) computed using the in-house code and the {\tt ggxy} package at different center-of-mass energies. The uncertainties shown correspond to 7-point scale variations.}
	}
	\label{tab:code_comparison_tot}
	\end{center}
\end{table}
\begin{table}[!htbp]
	\begin{center}
	\setlength{\extrarowheight}{6pt}
	\scalebox{0.70}{%
	\begin{tabular}{|c|c|c|c|c|}
		\hline
		$Q$ (GeV) & \multicolumn{2}{|c|}{\textbf{no Z-rad}} & \multicolumn{2}{|c|}{\textbf{with Z-rad}} \\
		\hline
		& In-house (fb) & {\tt ggxy} (fb) & In-house (fb) & {\tt ggxy} (fb) \\
		\hline
		
		250.00 
		& $1.286(2)\times10^{-1}$ 
		& $1.285(3)\times10^{-1}$ 
		& $1.298(2)\times10^{-1}$ 
		& $1.301(3)\times10^{-1}$ \\
		\hline
		
		350.00 
		& $1.118(3)$ 
		& $1.120(1)$ 
		& $1.119(3)$ 
		& $1.121(1)$ \\
		\hline
		
		800.00 
		& $5.543(2)\times10^{-3}$ 
		& $5.536(7)\times10^{-3}$ 
		& $5.696(2)\times10^{-3}$ 
		& $5.678(3)\times10^{-3}$ \\
		\hline
		
		1100.00 
		& $4.842(2)\times10^{-4}$ 
		& $4.831(1)\times10^{-4}$ 
		& $5.349(2)\times10^{-4}$ 
		& $5.342(1)\times10^{-4}$ \\
		\hline
		
		1400.00 
		& $7.152(3)\times10^{-5}$ 
		& $7.188(2)\times10^{-5}$ 
		& $9.094(3)\times10^{-5}$ 
		& $9.093(6)\times10^{-5}$ \\
		\hline
		
		1800.00 
		& $9.508(4)\times10^{-6}$ 
		& $9.510(6)\times10^{-6}$ 
		& $1.572(5)\times10^{-5}$ 
		& $1.586(6)\times10^{-5}$ \\
		\hline
	\end{tabular}
       }
	\caption{
		\small{Comparison of invariant mass distributions at NLO between the in-house code and the \texttt{ggxy} package. The Monte Carlo errors have been quoted along with the central scale value.
	}}
	\label{tab:code_comparison_inv}
	\end{center}
\end{table}
%
In figure~\ref{fig:InvMass_wBIT} we present the invariant-mass distribution for the exact NLO cross-section, and for the born improved NLO cross section.
In lower panel the respective K-factors are shown, where the born-improved K-factor, $K_{10}^{\ast}$ is seen to completely miss the peak around $Q=2m_{t}$ GeV, due to lack of top mass effects in the two-loop virtual amplitudes. 
%
In the region $ 2 m_{t}< Q < 2500$ GeV, the born-improved results mildly decrease, whereas the exact theory results first decrease sharply and then increase. Beyond 2500 GeV, both the results starts approaching each other.
%
\begin{figure}[ht]
\centering
\includegraphics[width=0.49\textwidth]{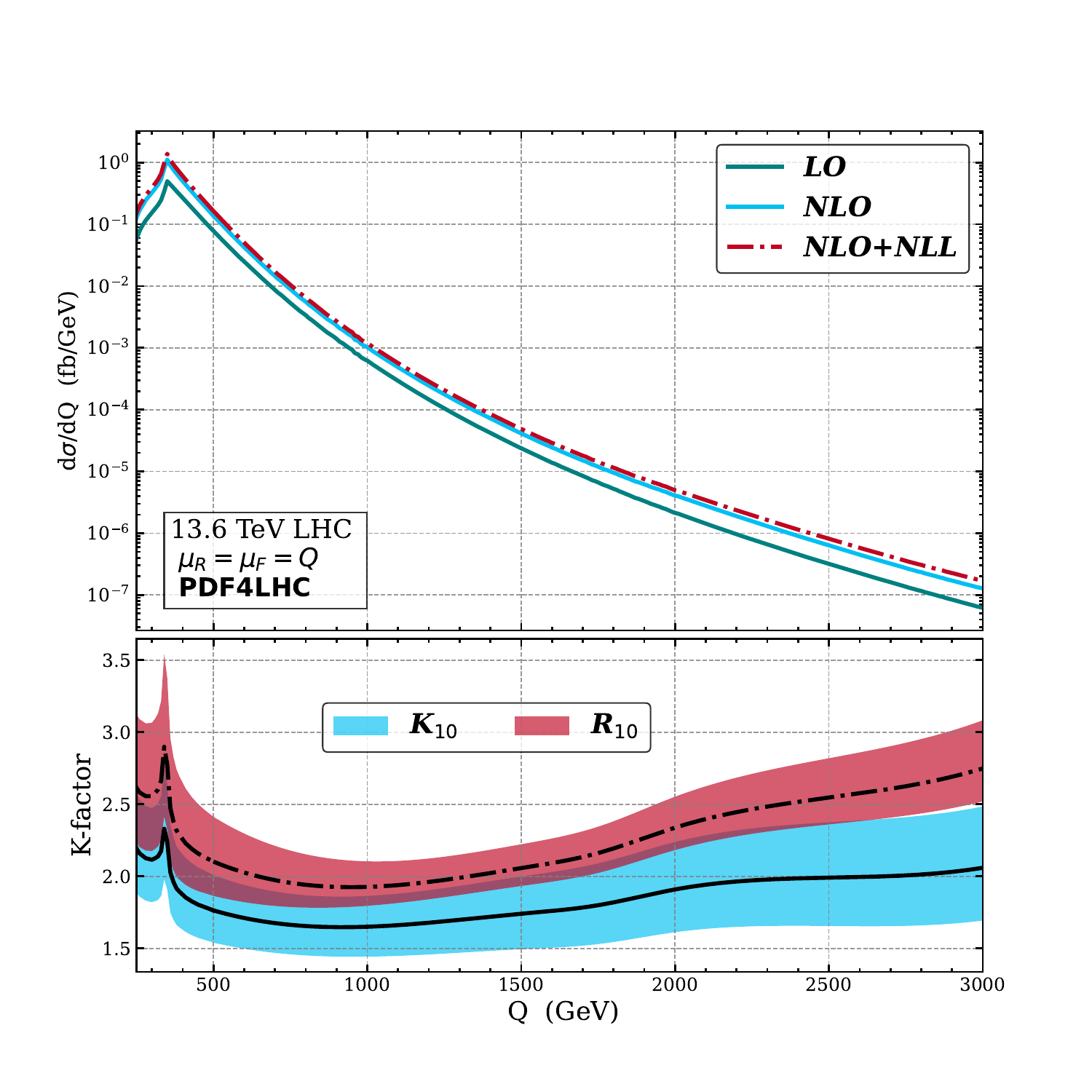}
\includegraphics[width=0.49\textwidth]{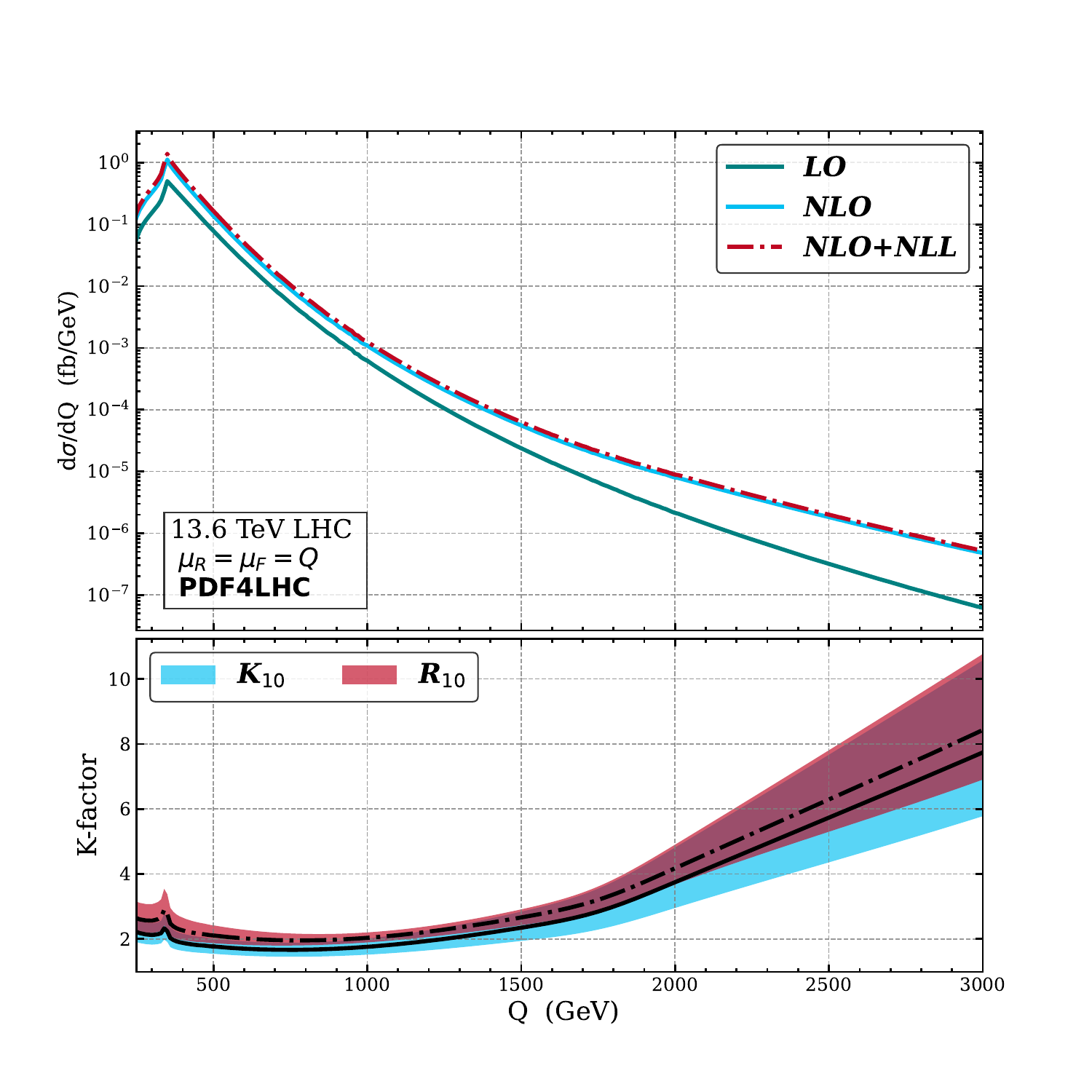}
\caption{
\small{
        Invariant mass distribution for the gluon fusion channel at NLO with(right) and without the $Z$-radiated diagrams(left). The respective K-factors are shown in the lower panel of the figure.
        }
        }
\label{fig:InvMass_ZandnoZrad}
\end{figure}
In figure~\ref{fig:InvMass_ZandnoZrad} we present the fixed-order as well the resummed cross-section along with the respective K-factors ($K_{10}$ and $R_{10}$). 
The left panel shows the cross-sections without including the $Z$-radiated type of diagrams, whereas the right panel shows the full cross-section, including the $Z$-radiated diagrams. 
The band in the lower panel of the figures are the uncertainties from the 7 point scale variation. 
While  $K_{10}$ for $Z$-radiated diagrams goes as high as $7.74$ at the central scale choice for $Q=3000$ GeV, it is around $2.06$ for the non-$Z$ radiated counterpart.
After the resummation of threshold logarithms to all orders, for the central scale choice, the $R_{10}$  is $8.42$ ($2.74$) for the $Z$-radiated diagrams (no $Z$-radiated) at $Q=3000$ GeV.
In general the band width of the $K_{10}$ is found to increase with $Q$ and is estimated to be about $4.79$ ($0.79$) for the $Z$-radiated (no $Z$-radiated) case at $Q=3000$GeV. 
After resummation the corresponding band width gets reduced to $3.86$ ($0.56$).
To better see this behaviour we present the scale uncertainties in the fixed-order as well as in the resummed results for different $Q$-values in figure~\ref{fig:7pt_unc}.
The scale uncertainty is quantified as,
\begin{align}
	\mu = \mathrm{max}\left| \frac{\sigma_{\rm NLO} - \sigma_{\rm NLO,c}}{\sigma_{\rm NLO,c}} \right| \times 100 \%
\end{align}
We find that the 7-point scale uncertainties $\mu_R=\mu_F=\mu$ for the NLO cross-secition is about $18.7\%$ at $Q=2m_{t}$, which goes down to a minimum of $12.8\%$ at $Q=850$GeV, and then increases in the high-$Q$  region to about $20.7\%$ for $Q=3000$ GeV. 
After performing the resummation, we find that in the low-$Q$ ($Q=2 m_{t}$) region the scale uncertainty increase to about $22.2\%$ and the start decreasing to a minimum of about $8 \%$ at $Q=1500$ GeV. 
Further as we move to high-$Q$ region the scale uncertainties in the resummed prediction increase slowly to about $12.21\%$ at $Q=3000$ GeV. 
\begin{figure}[!htbp]
\centering
\includegraphics[width=0.49\textwidth]{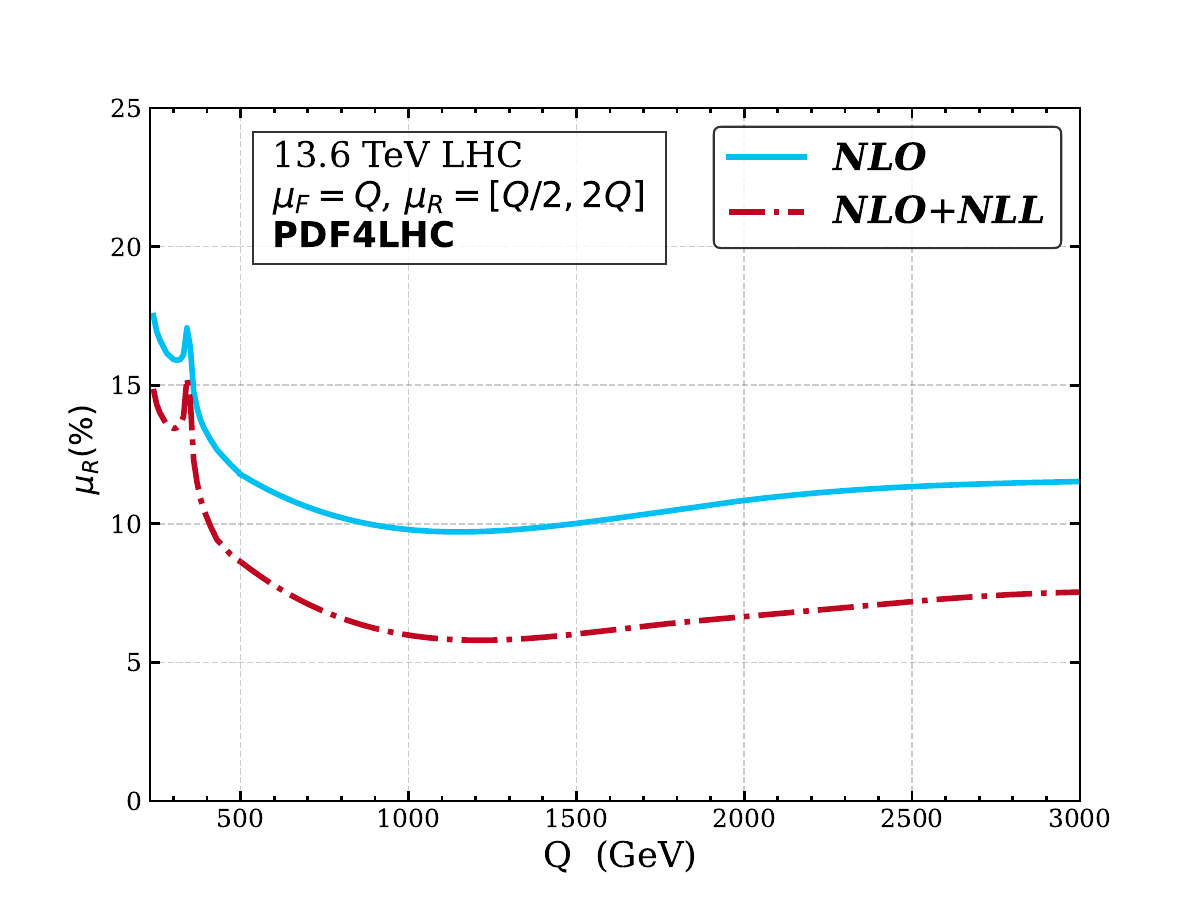}
\includegraphics[width=0.49\textwidth]{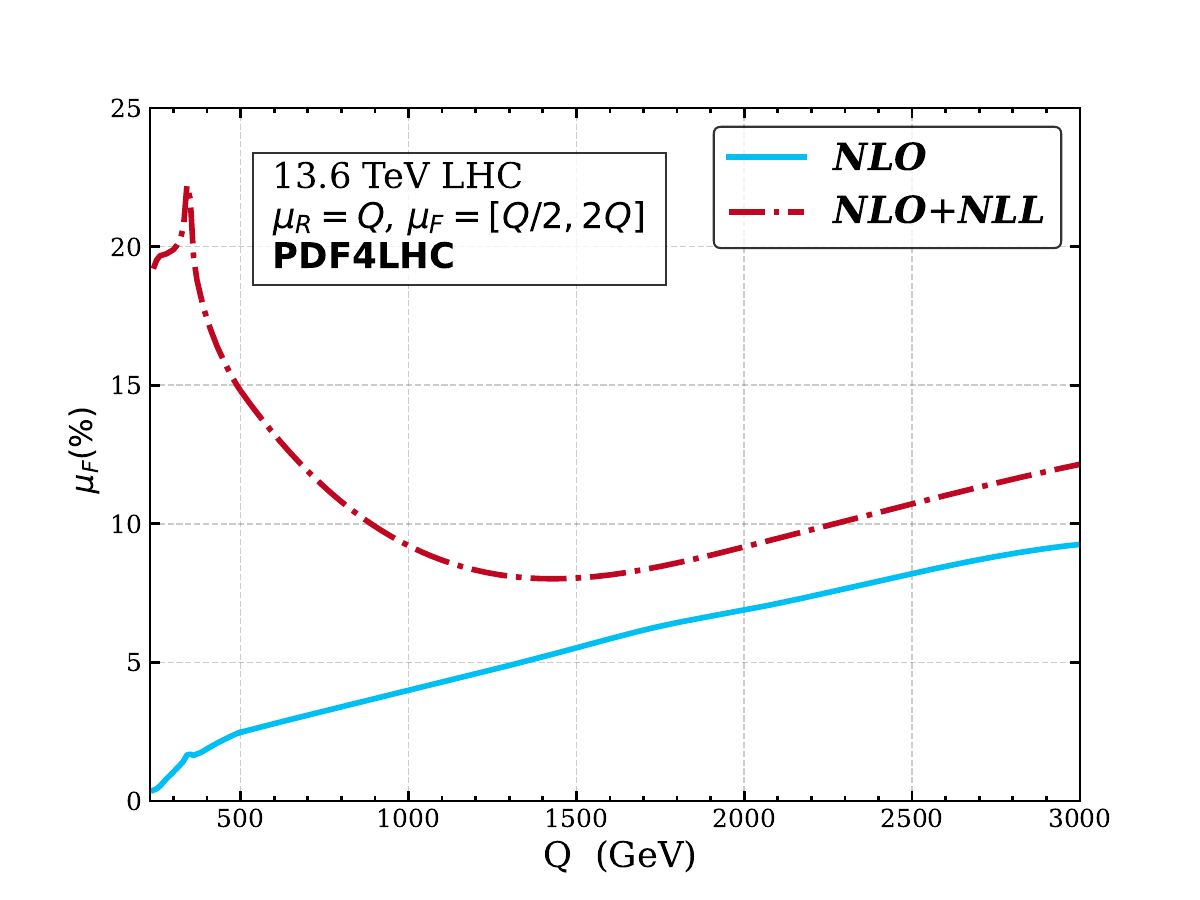}
\caption{
\small{
        The percentage uncertainty from the renormalization scale variation (left) and factorization scale variation (right) for the invariant mass distribution without the Z-radiated diagrams at NLO and NLO$+$NLL accuracy.
}
}\label{fig:scale_unc_mur_muf}
\end{figure}
%
It is worthwhile to keep one of the unphysical scales fixed to the central scale choice, while varying the other one from $Q/2$ and $2Q$. This is represented in figure~\ref{fig:scale_unc_mur_muf}. We observe that due to threshold resummation, the renormalization scale uncertainties gets reduced throughout the invariant-mass range, whereas for the factorization scale uncertainties it gets enhanced. The reduction in the renormalization sacle uncertainties is more in the high-$Q$ region in comparison to the low $Q$-region. For exmaple, at $Q=2m_{t}$ the reduction in renormalization scale uncertainty is around $1.79\%$, whereas at $Q=3000$ GeV region it is around $3.99\%$.
This behavior for other color-singlet processes have been noted before in the literature ~\cite{Banerjee:2025tbo,Banerjee:2024cpr,Das:2022zie,AH:2020cok}

\begin{figure}[!htbp]
\centering
\includegraphics[width=0.6\textwidth]{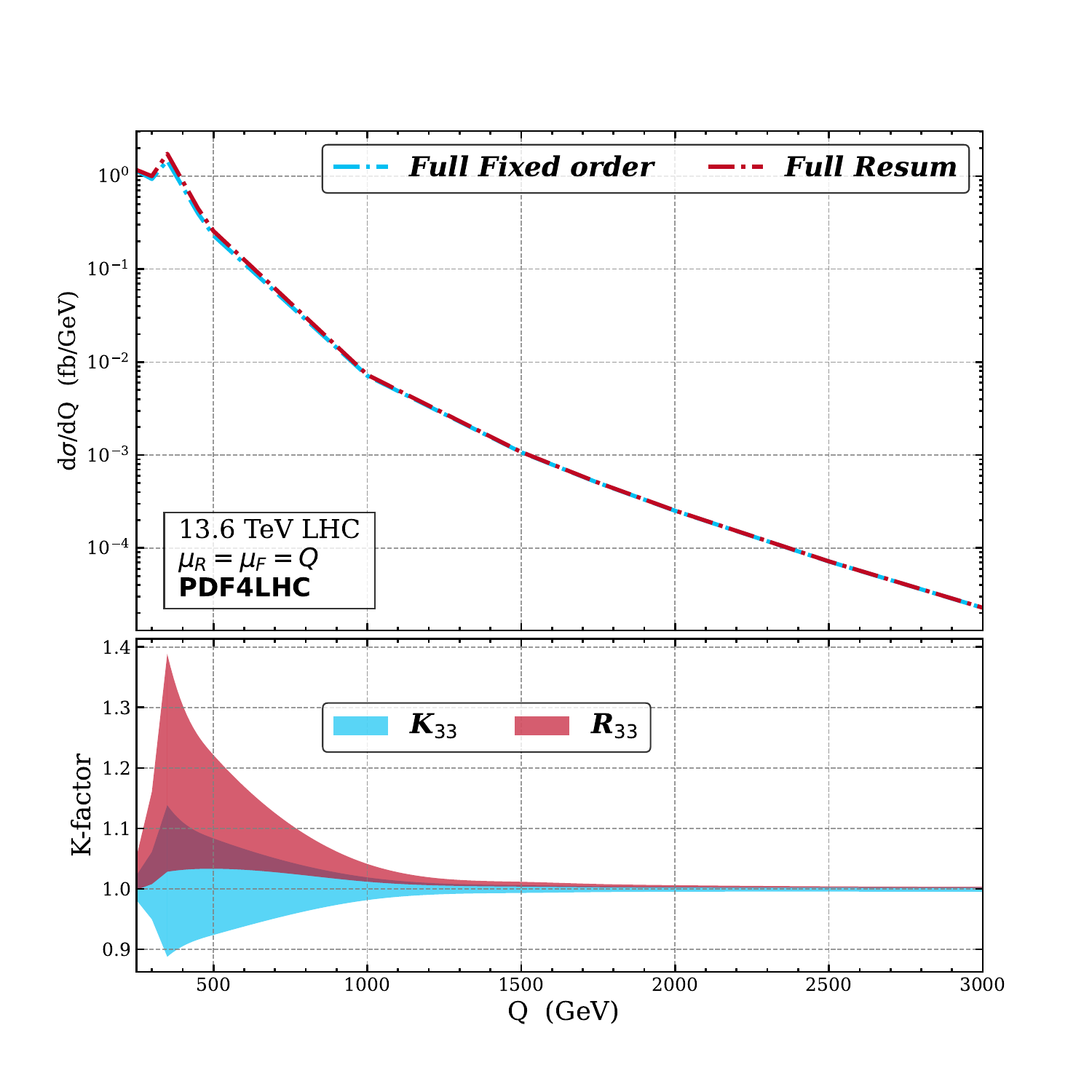}
\caption{
\small{
        Fixed order(resummed) Invariant mass distribution for $p p \to Z H$, including both the Drell-Yan at N$^{3}$LO (N$^3$LO+N$^3$LL) type~\cite{Das:2025wbj} and gluon fusion contributions at NLO (NLO+NLL). The respective K-factors are shown in the lower panel of the figure.
        }
}\label{fig:Invmass_full}
\end{figure}

In figure~\ref{fig:Invmass_full} we present the invariant mass distribution for the full $ZH$ production process, including both the Drell-Yan type and gluon fusion contributions at NLO with the exact two-loop virtual amplitude for the $gg$ channel. 
The band in lower panel of the figure shows the 7-point scale uncertainties.
The band width is presented with respect to the full fixed order results at central scale choice.
The scale uncertainties in the resummed results are larger in the low-Q region (around $Q = 2m_{t}$) than those in the high-Q region ($Q>2000$ GeV). Similar behaviour can be seen in the fixed order results as well.
\begin{table}[h!]
\begin{center}
\setlength{\extrarowheight}{7pt}
\scalebox{0.9}{%
\begin{tabular}{|c|c|c|c|c|c|}
                                \hline
                                       & 8 TeV  & 13 TeV  & 13.6 TeV  &  14 TeV & 100 TeV \\
                                \hline
                                \hline
                                $\sigma^{\text{NLO}}_{\text{BIT}}$
                                &  $35.19 ^{+18.1\%}_{-15.4\%}$  &  $112.81 ^{+15.5\%}_{-13.4\%}$ & $124.74 ^{+15.3\%}_{-13.2\%}$ & $132.99 ^{+15.3\%}_{-13.2\%}$ & $4311.03 ^{+14.9\%}_{-12.2\%}$ \\
                                \hline
                                $\sigma^{\text {NLO}}_{gg}$
                                &  $34.18 ^{+18.6\%}_{-15.5\%}$  &  $108.5 ^{+16.2\%}_{-13.7\%}$ & $119.82 ^{+16.0\%}_{-13.6\%}$ & $127.64 ^{+15.9\%}_{-13.5\%}$ & $4086.63 ^{+13.1\%}_{-11.2\%}$ \\
                                \hline
                                $\sigma^{\text {NLO+NLL}}_{gg}$
                                &  $42.63 ^{+18.1\%}_{-13.5\%}$  &  $130.95 ^{+19.5\%}_{-15.6\%}$ & $144.95 ^{+18.3\%}_{-13.8\%}$ & $153.62 ^{+19.1\%}_{-15.3\%}$ & $4665.56 ^{+20.9\%}_{-16.1\%}$ \\
                                \hline
                        \end{tabular}
                                }
                                \caption{
                                        \small{
                                                The $ZH$ production cross-sections (in fb) are presented at NLO with Born improved theory(BIT)
                                and NLO exact, along with corresponding SV resummed results at various hadronic entre of masses for the LHC with 7-point scale uncertainty.
                                }
                                }
                                \label{tab:tableggZH_total}
                \end{center}
\end{table}

In \tab{tab:tableggZH_total} we give the total cross-sections for the gluon fusion channel at fixed order as well as after  resummation, by integrating the invariant mass distributions up to $\sqrt{S}$. 
Alongside, we also provide uncertainty arising from $7$-point scale variation.
As also observed earlier for differential cross-sections, the born improved theory overestimates the full cross section. After resummation, the cross-section increases by about $20-24\%$, across different center of mass energies. 
For the total cross-section, the resummation at NLO+NLL improves the cross-section by around $20\%$ at 8 TeV and around $12\%$ at 100 TeV compared to the fixed order, whereas there is not much change in scale uncertainty. 
At 13.6 TeV, the exact NLO total production cross-section is about $4.1\%$ larger than the Born-improved theory (BIT) result. 
Including NLO+NLL resummation enhances the total production cross-section by about $21.0\%$ over the FO results. 
We note that the scale uncentainty increases after performing the resummation for all the center of mass energies taken here. 
This can be traced to the fact that, total cross-section receives significant contributions from the low invariant mass
regions, where the factorisation scale uncertainty remains significant.

For the resummation case, we improve the fixed order for the DY-type channel by SV resummation at N$^3$LO+N$^3$LL and for the
gluon channel by resummation at NLO+NLL with exact NLO. For the DY-type channel, resummation effect is very small at this order and
the major improvement is obtained through the gluon fusion channel. 
%

\section{Conclusion}\label{sec:conclusion}

We have presented the NLO+NLL predictions for the first time
for associated $ZH$ production through the gluon fusion channel with exact top-quark mass dependence in the virtual amplitudes,
thanks to {\tt ggxy} package which made this computation possible. 
The fixed-order NLO results were computed using our in-house developed code based on the dipole subtraction method. 
We studied both invariant-mass distributions and total production cross-sections over a wide range of collider energies and investigated the impact of threshold resummation on the perturbative predictions.
As the exact NLO prediction captures the relevant top mass effects, the results reported in this draft  give better predicitons than those in the Born-improved theory specifically near the $2 m_{t}$ region.
The contribution from $Z$-radiated diagrams was also analyzed separately, and these diagrams were found to generate significantly larger K-factors in the large invariant-mass region.

The inclusion of threshold logarithms at NLO+NLL accuracy enhances the gluon-fusion cross-section substantially. At $\sqrt{S}=13.6$ TeV, the total production cross-section increases by approximately $21\%$ relative to the fixed-order NLO result. For differential distributions, the resummed predictions improve the perturbative stability in the high-$Q$ region by reducing the seven-point scale uncertainty from about $20\%$ at NLO to nearly $12\%$ at $Q=3000$ GeV. The reduction in uncertainty is mainly driven by improved renormalization-scale dependence after resummation.

Finally, combining the Drell--Yan contribution at N$^3$LO+N$^3$LL accuracy with the gluon-fusion contribution at NLO+NLL accuracy provides the most precise QCD prediction for the  $ZH$ invariant mass distribution.

\raggedbottom
\enlargethispage{3\baselineskip}
\section*{Acknowledgements}
The work of C.D. at Physical Research Laboratory was supported by Department of Space, Govt. of India. The authors would like to thank G. Das, K. Samanta and V. Ravindran for the useful discussions.
We acknowledge National Supercomputing
Mission (NSM) for providing computing resources of ‘$\tt{PARAM \ Kamrupa}$’ at IIT Guwahati,
which is implemented by C-DAC and supported by the Ministry of Electronics and Informa-
tion Technology (MeitY) and Department of Science and Technology (DST), Government
of India, where most of the computational work has been carried out.



\begin{appendix}
\numberwithin{equation}{section}


\end{appendix}

\pagebreak





\bibliographystyle{JHEP}
\bibliography{gg2zh}

@article{ATLAS:2018kot,
    author = "Aaboud, Morad and others",
    collaboration = "ATLAS",
    title = "{Observation of $H \rightarrow b\bar{b}$ decays and $VH$ production with the ATLAS detector}",
    eprint = "1808.08238",
    archivePrefix = "arXiv",
    primaryClass = "hep-ex",
    reportNumber = "CERN-EP-2018-215",
    doi = "10.1016/j.physletb.2018.09.013",
    journal = "Phys. Lett. B",
    volume = "786",
    pages = "59--86",
    year = "2018"
}

@article{CMS:2018nsn,
    author = "Sirunyan, A. M. and others",
    collaboration = "CMS",
    title = "{Observation of Higgs boson decay to bottom quarks}",
    eprint = "1808.08242",
    archivePrefix = "arXiv",
    primaryClass = "hep-ex",
    reportNumber = "CMS-HIG-18-016, CERN-EP-2018-223",
    doi = "10.1103/PhysRevLett.121.121801",
    journal = "Phys. Rev. Lett.",
    volume = "121",
    number = "12",
    pages = "121801",
    year = "2018"
}

@article{Han:1991ia,
    author = "Han, Tao and Willenbrock, S.",
    title = "{QCD Correction to the $pp \to WH$ and $ZH$ Total Cross Sections}",
    reportNumber = "FERMILAB-PUB-91-070-T, BNL-45990",
    doi = "10.1016/0370-2693(91)90572-8",
    journal = "Phys. Lett. B",
    volume = "273",
    pages = "167--172",
    year = "1991"
}

@article{Brein:2003wg,
    author = "Brein, Oliver and Djouadi, Abdelhak and Harlander, Robert",
    title = "{NNLO QCD corrections to the Higgs-strahlung processes at hadron colliders}",
    eprint = "hep-ph/0307206",
    archivePrefix = "arXiv",
    reportNumber = "MPP-2003-35, CERN-TH-2003-161, PM-03-16",
    doi = "10.1016/j.physletb.2003.10.112",
    journal = "Phys. Lett. B",
    volume = "579",
    pages = "149--156",
    year = "2004"
}

@article{Baglio:2022wzu,
    author = "Baglio, Julien and Duhr, Claude and Mistlberger, Bernhard and Szafron, Robert",
    title = "{Inclusive production cross sections at N$^{3}$LO}",
    eprint = "2209.06138",
    archivePrefix = "arXiv",
    primaryClass = "hep-ph",
    reportNumber = "CERN-TH-2022-109, SLAC-PUB-17699, BONN-TH-2022-22",
    doi = "10.1007/JHEP12(2022)066",
    journal = "JHEP",
    volume = "12",
    pages = "066",
    year = "2022"
}

@article{Dicus:1988yh,
    author = "Dicus, Duane A. and Kao, Chung",
    title = "{Higgs Boson - $Z^0$ Production From Gluon Fusion}",
    reportNumber = "DOE-ER40200-131",
    doi = "10.1103/PhysRevD.38.1008",
    journal = "Phys. Rev. D",
    volume = "38",
    pages = "1008",
    year = "1988",
    note = "[Erratum: Phys.Rev.D 42, 2412 (1990)]"
}

@article{Kniehl:1990iva,
    author = "Kniehl, Bernd A.",
    title = "{Associated Production of Higgs and Z Bosons From Gluon Fusion in Hadron Collisions}",
    reportNumber = "MAD-PH-558",
    doi = "10.1103/PhysRevD.42.2253",
    journal = "Phys. Rev. D",
    volume = "42",
    pages = "2253--2258",
    year = "1990"
}

@article{Ciccolini:2003jy,
    author = "Ciccolini, M. L. and Dittmaier, S. and Kramer, M.",
    title = "{Electroweak radiative corrections to associated WH and ZH production at hadron colliders}",
    eprint = "hep-ph/0306234",
    archivePrefix = "arXiv",
    reportNumber = "EDINBURGH-2003-05, MPI-PHT-2003-24",
    doi = "10.1103/PhysRevD.68.073003",
    journal = "Phys. Rev. D",
    volume = "68",
    pages = "073003",
    year = "2003"
}

@article{Denner:2011id,
    author = "Denner, Ansgar and Dittmaier, Stefan and Kallweit, Stefan and Muck, Alexander",
    title = "{Electroweak corrections to Higgs-strahlung off W/Z bosons at the Tevatron and the LHC with HAWK}",
    eprint = "1112.5142",
    archivePrefix = "arXiv",
    primaryClass = "hep-ph",
    reportNumber = "FR-PHENO-2011-025, PSI-PR-11-04, ZU-TH-29-11, TTK-11-61",
    doi = "10.1007/JHEP03(2012)075",
    journal = "JHEP",
    volume = "03",
    pages = "075",
    year = "2012"
}

@article{Denner:2014cla,
    author = {Denner, Ansgar and Dittmaier, Stefan and Kallweit, Stefan and M{\"u}ck, Alexander},
    title = "{HAWK  2.0: A Monte Carlo program for Higgs production in vector-boson fusion and Higgs strahlung at hadron colliders}",
    eprint = "1412.5390",
    archivePrefix = "arXiv",
    primaryClass = "hep-ph",
    reportNumber = "FR-PHENO-2014-013, MITP-14-101, TTK-14-36",
    doi = "10.1016/j.cpc.2015.04.021",
    journal = "Comput. Phys. Commun.",
    volume = "195",
    pages = "161--171",
    year = "2015"
}

@article{ATLAS:2020fcp,
    author = "Aad, Georges and others",
    collaboration = "ATLAS",
    title = "{Measurements of $WH$ and $ZH$ production in the $H \rightarrow b\bar{b}$ decay channel in $pp$ collisions at 13 TeV with the ATLAS detector}",
    eprint = "2007.02873",
    archivePrefix = "arXiv",
    primaryClass = "hep-ex",
    reportNumber = "CERN-EP-2020-087",
    doi = "10.1140/epjc/s10052-020-08677-2",
    journal = "Eur. Phys. J. C",
    volume = "81",
    number = "2",
    pages = "178",
    year = "2021"
}

@article{ATLAS:2019yhn,
    author = "Aaboud, Morad and others",
    collaboration = "ATLAS",
    title = "{Measurement of VH, $ \mathrm{H}\to \mathrm{b}\overline{\mathrm{b}} $ production as a function of the vector-boson transverse momentum in 13 TeV pp collisions with the ATLAS detector}",
    eprint = "1903.04618",
    archivePrefix = "arXiv",
    primaryClass = "hep-ex",
    reportNumber = "CERN-EP-2019-019",
    doi = "10.1007/JHEP05(2019)141",
    journal = "JHEP",
    volume = "05",
    pages = "141",
    year = "2019"
}

@article{ATLAS:2020jwz,
    author = "Aad, Georges and others",
    collaboration = "ATLAS",
    title = "{Measurement of the associated production of a Higgs boson decaying into $b$-quarks with a vector boson at high transverse momentum in $pp$ collisions at $\sqrt{s} = 13$ TeV with the ATLAS detector}",
    eprint = "2008.02508",
    archivePrefix = "arXiv",
    primaryClass = "hep-ex",
    reportNumber = "CERN-EP-2020-093",
    doi = "10.1016/j.physletb.2021.136204",
    journal = "Phys. Lett. B",
    volume = "816",
    pages = "136204",
    year = "2021"
}

@article{Kumar:2014uwa,
    author = "Kumar, M. C. and Mandal, M. K. and Ravindran, V.",
    title = "{Associated production of Higgs boson with vector boson at threshold N$^{3}$LO in QCD}",
    eprint = "1412.3357",
    archivePrefix = "arXiv",
    primaryClass = "hep-ph",
    reportNumber = "HRI-RECAPP-2014-027",
    doi = "10.1007/JHEP03(2015)037",
    journal = "JHEP",
    volume = "03",
    pages = "037",
    year = "2015"
}

@article{Das:2022zie,
    author = "Das, Goutam and Dey, Chinmoy and Kumar, M. C. and Samanta, Kajal",
    title = "{Threshold enhanced cross sections for colorless productions}",
    eprint = "2210.17534",
    archivePrefix = "arXiv",
    primaryClass = "hep-ph",
    reportNumber = "TTK-22-34, P3H-22-106",
    doi = "10.1103/PhysRevD.107.034038",
    journal = "Phys. Rev. D",
    volume = "107",
    number = "3",
    pages = "034038",
    year = "2023"
}

@article{Harlander:2018yio,
    author = "Harlander, Robert V. and Klappert, Jonas and Liebler, Stefan and Simon, Lukas",
    title = "{vh@nnlo-v2: New physics in Higgs Strahlung}",
    eprint = "1802.04817",
    archivePrefix = "arXiv",
    primaryClass = "hep-ph",
    reportNumber = "KA-TP-01-2018, TTK-17-47",
    doi = "10.1007/JHEP05(2018)089",
    journal = "JHEP",
    volume = "05",
    pages = "089",
    year = "2018"
}

@article{Altenkamp:2012sx,
    author = "Altenkamp, Lukas and Dittmaier, Stefan and Harlander, Robert V. and Rzehak, Heidi and Zirke, Tom J. E.",
    title = "{Gluon-induced Higgs-strahlung at next-to-leading order QCD}",
    eprint = "1211.5015",
    archivePrefix = "arXiv",
    primaryClass = "hep-ph",
    reportNumber = "CERN-PH-TH-2012-312, FR-PHENO-2012-023, WUB-12-21",
    doi = "10.1007/JHEP02(2013)078",
    journal = "JHEP",
    volume = "02",
    pages = "078",
    year = "2013"
}

@article{Bizon:2021rww,
    author = {Bizo{\'n}, Wojciech and Caola, Fabrizio and Melnikov, Kirill and R{\"o}ntsch, Raoul},
    title = "{Anomalous couplings in associated VH production with Higgs boson decay to massive b quarks at NNLO in QCD}",
    eprint = "2106.06328",
    archivePrefix = "arXiv",
    primaryClass = "hep-ph",
    doi = "10.1103/PhysRevD.105.014023",
    journal = "Phys. Rev. D",
    volume = "105",
    number = "1",
    pages = "014023",
    year = "2022"
}

@article{Bizon:2019tfo,
    author = "Bizo{\'n}, Wojciech and Re, Emanuele and Zanderighi, Giulia",
    title = "{NNLOPS description of the $H \to b\overline{b} $ decay with MiNLO}",
    eprint = "1912.09982",
    archivePrefix = "arXiv",
    primaryClass = "hep-ph",
    reportNumber = "TTP19-049, LAPTH-052/19, MPP-2019-263",
    doi = "10.1007/JHEP06(2020)006",
    journal = "JHEP",
    volume = "06",
    pages = "006",
    year = "2020"
}

@article{Alioli:2019qzz,
    author = "Alioli, Simone and Broggio, Alessandro and Kallweit, Stefan and Lim, Matthew A. and Rottoli, Luca",
    title = "{Higgsstrahlung at NNLL'$+$NNLO matched to parton showers in GENEVA}",
    eprint = "1909.02026",
    archivePrefix = "arXiv",
    primaryClass = "hep-ph",
    doi = "10.1103/PhysRevD.100.096016",
    journal = "Phys. Rev. D",
    volume = "100",
    number = "9",
    pages = "096016",
    year = "2019"
}

@article{Harlander:2013mla,
    author = "Harlander, Robert V. and Liebler, Stefan and Zirke, Tom",
    title = "{Higgs Strahlung at the Large Hadron Collider in the 2-Higgs-Doublet Model}",
    eprint = "1307.8122",
    archivePrefix = "arXiv",
    primaryClass = "hep-ph",
    reportNumber = "WUB-13-12",
    doi = "10.1007/JHEP02(2014)023",
    journal = "JHEP",
    volume = "02",
    pages = "023",
    year = "2014"
}

@article{Englert:2013vua,
    author = "Englert, Christoph and McCullough, Matthew and Spannowsky, Michael",
    title = "{Gluon-initiated associated production boosts Higgs physics}",
    eprint = "1310.4828",
    archivePrefix = "arXiv",
    primaryClass = "hep-ph",
    reportNumber = "DCPT-13-158, IPPP-13-79, MIT-CTP-4506",
    doi = "10.1103/PhysRevD.89.013013",
    journal = "Phys. Rev. D",
    volume = "89",
    number = "1",
    pages = "013013",
    year = "2014"
}

@article{Gauld:2021ule,
    author = "Gauld, R. and Gehrmann-De Ridder, A. and Glover, E. W. N. and Huss, A. and Majer, I.",
    title = "{VH + jet production in hadron-hadron collisions up to order $ {\alpha}_{\mathrm{s}}^3 $ in perturbative QCD}",
    eprint = "2110.12992",
    archivePrefix = "arXiv",
    primaryClass = "hep-ph",
    reportNumber = "NIKHEF 2021-026, BONN-TH-2021-09, IPPP/21/26, ZU-TH 49/21,
  CERN-TH-2021-159",
    doi = "10.1007/JHEP03(2022)008",
    journal = "JHEP",
    volume = "03",
    pages = "008",
    year = "2022"
}

@article{Englert:2016hvy,
    author = "Englert, Christoph and Rosenfeld, Rogerio and Spannowsky, Michael and Tonero, Alberto",
    title = "{New physics and signal-background interference in associated $pp\to HZ$ production}",
    eprint = "1603.05304",
    archivePrefix = "arXiv",
    primaryClass = "hep-ph",
    reportNumber = "IPPP-16-21, DCPT-16-42",
    doi = "10.1209/0295-5075/114/31001",
    journal = "EPL",
    volume = "114",
    number = "3",
    pages = "31001",
    year = "2016"
}

@article{LHCHiggsCrossSectionWorkingGroup:2016ypw,
    author = "de Florian, D. and others",
    collaboration = "LHC Higgs Cross Section Working Group",
    title = "{Handbook of LHC Higgs Cross Sections: 4. Deciphering the Nature of the Higgs Sector}",
    eprint = "1610.07922",
    archivePrefix = "arXiv",
    primaryClass = "hep-ph",
    reportNumber = "CERN-2017-002-M, CERN-2017-002",
    doi = "10.23731/CYRM-2017-002",
    journal = "CERN Yellow Rep. Monogr.",
    volume = "2",
    pages = "1--869",
    year = "2017"
}

@article{Harlander:2018yns,
    author = "Harlander, R. V. and Klappert, J. and Pandini, C. and Papaefstathiou, A.",
    title = "{Exploiting the WH/ZH symmetry in the search for New Physics}",
    eprint = "1804.02299",
    archivePrefix = "arXiv",
    primaryClass = "hep-ph",
    reportNumber = "TTK-17-48, Nikhef 2018-026",
    doi = "10.1140/epjc/s10052-018-6234-x",
    journal = "Eur. Phys. J. C",
    volume = "78",
    number = "9",
    pages = "760",
    year = "2018"
}

@article{Hasselhuhn:2016rqt,
    author = "Hasselhuhn, Alexander and Luthe, Thomas and Steinhauser, Matthias",
    title = "{On top quark mass effects to $gg\to ZH$ at NLO}",
    eprint = "1611.05881",
    archivePrefix = "arXiv",
    primaryClass = "hep-ph",
    reportNumber = "TTP16-051",
    doi = "10.1007/JHEP01(2017)073",
    journal = "JHEP",
    volume = "01",
    pages = "073",
    year = "2017"
}

@article{Davies:2025out,
    author = {Davies, Joshua and Grau, Dominik and Sch{\"o}nwald, Kay and Steinhauser, Matthias and Stremmer, Daniel and Vitti, Marco},
    title = "{Two-loop QCD corrections to ZH and off-shell Z boson pair production in gluon fusion}",
    eprint = "2509.07072",
    archivePrefix = "arXiv",
    primaryClass = "hep-ph",
    reportNumber = "P3H-25-060, TTP25-029, ZU-TH 56/25",
    doi = "10.1007/JHEP03(2026)200",
    journal = "JHEP",
    volume = "03",
    pages = "200",
    year = "2026"
}

@article{Yan:2021veo,
    author = "Yan, Bin and Yuan, C. -P.",
    title = "{Anomalous Zbb{\textasciimacron} Couplings: From LEP to LHC}",
    eprint = "2101.06261",
    archivePrefix = "arXiv",
    primaryClass = "hep-ph",
    reportNumber = "MSUHEP-21-001,LA-UR-21-20373, MSUHEP-21-001",
    doi = "10.1103/PhysRevLett.127.051801",
    journal = "Phys. Rev. Lett.",
    volume = "127",
    number = "5",
    pages = "051801",
    year = "2021"
}

@article{Ferrera:2014lca,
    author = "Ferrera, Giancarlo and Grazzini, Massimiliano and Tramontano, Francesco",
    title = "{Associated ZH production at hadron colliders: the fully differential NNLO QCD calculation}",
    eprint = "1407.4747",
    archivePrefix = "arXiv",
    primaryClass = "hep-ph",
    reportNumber = "IFUM-1031-FT, ZU-TH-23-14",
    doi = "10.1016/j.physletb.2014.11.040",
    journal = "Phys. Lett. B",
    volume = "740",
    pages = "51--55",
    year = "2015"
}

@article{Hespel:2015zea,
    author = "Hespel, B. and Maltoni, F. and Vryonidou, E.",
    title = "{Higgs and Z boson associated production via gluon fusion in the SM and the 2HDM}",
    eprint = "1503.01656",
    archivePrefix = "arXiv",
    primaryClass = "hep-ph",
    reportNumber = "CP3-15-05, MCNET-15-04",
    doi = "10.1007/JHEP06(2015)065",
    journal = "JHEP",
    volume = "06",
    pages = "065",
    year = "2015"
}

@article{Astill:2018ivh,
    author = "Astill, William and Bizo{\'n}, Wojciech and Re, Emanuele and Zanderighi, Giulia",
    title = "{NNLOPS accurate associated HZ production with $ H\to b\overline{b} $ decay at NLO}",
    eprint = "1804.08141",
    archivePrefix = "arXiv",
    primaryClass = "hep-ph",
    reportNumber = "CERN-TH-2018-082, LAPTH-014/18, OUTP-17-18P, LAPTH-014-18",
    doi = "10.1007/JHEP11(2018)157",
    journal = "JHEP",
    volume = "11",
    pages = "157",
    year = "2018"
}

@article{Chen:2020gae,
    author = "Chen, Long and Heinrich, Gudrun and Jones, Stephen P. and Kerner, Matthias and Klappert, Jonas and Schlenk, Johannes",
    title = "{$ZH$ production in gluon fusion: two-loop amplitudes with full top quark mass dependence}",
    eprint = "2011.12325",
    archivePrefix = "arXiv",
    primaryClass = "hep-ph",
    reportNumber = "ZU-TH 45/20, CERN-TH-2020-199, IPPP/20/57, P3H-20-076,
  KA-TP-21-2020, P3H-20-076, KA-TP-21-2020, TTK-20-42, PSI-PR-20-21, PSI-PR-20-21 P3H-20-076{\textbackslash}{\textbackslash} KA-TP-21-2020{\textbackslash}{\textbackslash} TTK-20-42{\textbackslash}{\textbackslash} PSI-PR-20-21",
    doi = "10.1007/JHEP03(2021)125",
    journal = "JHEP",
    volume = "03",
    pages = "125",
    year = "2021"
}

@article{Chen:2022rua,
    author = "Chen, Long and Davies, Joshua and Heinrich, Gudrun and Jones, Stephen P. and Kerner, Matthias and Mishima, Go and Schlenk, Johannes and Steinhauser, Matthias",
    title = "{ZH production in gluon fusion at NLO in QCD}",
    eprint = "2204.05225",
    archivePrefix = "arXiv",
    primaryClass = "hep-ph",
    reportNumber = "IPPP/22/19,P3H-22-038,KA-TP-08-2022,TTP22-024,TU-1147,PSI-PR-22-08",
    doi = "10.1007/JHEP08(2022)056",
    journal = "JHEP",
    volume = "08",
    pages = "056",
    year = "2022"
}

@article{Alasfar:2021ppe,
    author = {Alasfar, Lina and Degrassi, Giuseppe and Giardino, Pier Paolo and Gr{\"o}ber, Ramona and Vitti, Marco},
    title = "{Virtual corrections to $gg\to ZH$ via a transverse momentum expansion}",
    eprint = "2103.06225",
    archivePrefix = "arXiv",
    primaryClass = "hep-ph",
    doi = "10.1007/JHEP05(2021)168",
    journal = "JHEP",
    volume = "05",
    pages = "168",
    year = "2021"
}

@article{Degrassi:2022mro,
    author = {Degrassi, Giuseppe and Gr{\"o}ber, Ramona and Vitti, Marco and Zhao, Xiaoran},
    title = "{On the NLO QCD corrections to gluon-initiated ZH production}",
    eprint = "2205.02769",
    archivePrefix = "arXiv",
    primaryClass = "hep-ph",
    reportNumber = "CERN-TH-2022-079",
    doi = "10.1007/JHEP08(2022)009",
    journal = "JHEP",
    volume = "08",
    pages = "009",
    year = "2022"
}

@article{Davies:2025otz,
    author = {Davies, Joshua and Grau, Dominik and Sch{\"o}nwald, Kay and Steinhauser, Matthias and Stremmer, Daniel},
    title = "{Three-loop corrections to gg {\textrightarrow} ZH in the large top quark mass limit}",
    eprint = "2512.00156",
    archivePrefix = "arXiv",
    primaryClass = "hep-ph",
    reportNumber = "CERN-TH-2025-245, P3H-25-101, TTP25-047",
    doi = "10.1007/JHEP02(2026)191",
    journal = "JHEP",
    volume = "02",
    pages = "191",
    year = "2026"
}

@article{Wang:2021rxu,
    author = "Wang, Guoxing and Xu, Xiaofeng and Xu, Yongqi and Yang, Li Lin",
        title = "{Next-to-leading order corrections for $gg \to ZH$ with top quark mass dependence}",
    eprint = "2107.08206",
    archivePrefix = "arXiv",
    primaryClass = "hep-ph",
    doi = "10.1016/j.physletb.2022.137087",
    journal = "Phys. Lett. B",
    volume = "829",
    pages = "137087",
    year = "2022"
}

@article{Das:2025wbj,
    author = "Das, Goutam and Dey, Chinmoy and Kumar, M. C. and Samanta, Kajal",
    title = "{Soft gluon resummation for gluon fusion ZH production}",
    eprint = "2501.10330",
    archivePrefix = "arXiv",
    primaryClass = "hep-ph",
    reportNumber = "TTK-24-58, P3H-24-100, IPPP/24/81",
    doi = "10.1103/mxxw-bvm3",
    journal = "Phys. Rev. D",
    volume = "113",
    number = "1",
    pages = "014024",
    year = "2026"
}

@article{Ahmed:2019udm,
    author = "Ahmed, Taushif and Ajjath, A. H. and Chen, Long and Dhani, Prasanna K. and Mukherjee, Pooja and Ravindran, V.",
    title = "{Polarised Amplitudes and Soft-Virtual Cross Sections for $b\bar b \rightarrow ZH$ at NNLO in QCD}",
    eprint = "1910.06347",
    archivePrefix = "arXiv",
    primaryClass = "hep-ph",
    reportNumber = "IMSc/2019/10/09,MPP-2019-202",
    doi = "10.1007/JHEP01(2020)030",
    journal = "JHEP",
    volume = "01",
    pages = "030",
    year = "2020"
}

@article{Catani:2014uta,
    author = "Catani, Stefano and Cieri, Leandro and de Florian, Daniel and Ferrera, Giancarlo and Grazzini, Massimiliano",
    title = "{Threshold resummation at N$^3$LL accuracy and soft-virtual cross sections at N$^3$LO}",
    eprint = "1405.4827",
    archivePrefix = "arXiv",
    primaryClass = "hep-ph",
    reportNumber = "ZU-TH-20-14",
    doi = "10.1016/j.nuclphysb.2014.09.012",
    journal = "Nucl. Phys. B",
    volume = "888",
    pages = "75--91",
    year = "2014"
}

@article{Das:2019btv,
    author = "Das, Goutam and Moch, Sven-Olaf and Vogt, Andreas",
    title = "{Soft corrections to inclusive deep-inelastic scattering at four loops and beyond}",
    eprint = "1912.12920",
    archivePrefix = "arXiv",
    primaryClass = "hep-ph",
    reportNumber = "SI-HEP-2019-21, DESY 19-088, LTH 1205, DESY-19-088",
    doi = "10.1007/JHEP03(2020)116",
    journal = "JHEP",
    volume = "03",
    pages = "116",
    year = "2020"
}

@article{Das:2020adl,
    author = "Das, G. and Moch, S. and Vogt, A.",
    title = "{Approximate four-loop QCD corrections to the Higgs-boson production cross section}",
    eprint = "2004.00563",
    archivePrefix = "arXiv",
    primaryClass = "hep-ph",
    reportNumber = "SI-HEP-2020-07, DESY 20-037, DESY-20-037, LTH 1230",
    doi = "10.1016/j.physletb.2020.135546",
    journal = "Phys. Lett. B",
    volume = "807",
    pages = "135546",
    year = "2020"
}

@article{Sterman:1986aj,
    author = "Sterman, George F.",
    title = "{Summation of Large Corrections to Short Distance Hadronic Cross-Sections}",
    reportNumber = "Print-86-0873 (IAS,PRINCETON)",
    doi = "10.1016/0550-3213(87)90258-6",
    journal = "Nucl. Phys. B",
    volume = "281",
    pages = "310--364",
    year = "1987"
}

@article{Catani:1989ne,
    author = "Catani, S. and Trentadue, L.",
    title = "{Resummation of the QCD Perturbative Series for Hard Processes}",
    reportNumber = "DFF-93/3/89",
    doi = "10.1016/0550-3213(89)90273-3",
    journal = "Nucl. Phys. B",
    volume = "327",
    pages = "323--352",
    year = "1989"
}

@article{Catani:2003zt,
    author = "Catani, Stefano and de Florian, Daniel and Grazzini, Massimiliano and Nason, Paolo",
    title = "{Soft gluon resummation for Higgs boson production at hadron colliders}",
    eprint = "hep-ph/0306211",
    archivePrefix = "arXiv",
    reportNumber = "BICOCCA-FT-03-12, CERN-TH-2003-117",
    doi = "10.1088/1126-6708/2003/07/028",
    journal = "JHEP",
    volume = "07",
    pages = "028",
    year = "2003"
}

@article{Moch:2005ky,
    author = "Moch, S. and Vogt, A.",
    title = "{Higher-order soft corrections to lepton pair and Higgs boson production}",
    eprint = "hep-ph/0508265",
    archivePrefix = "arXiv",
    reportNumber = "DESY-05-152, SFB-CPP-05-46, DCPT-05-110, IPPP-05-55",
    doi = "10.1016/j.physletb.2005.09.061",
    journal = "Phys. Lett. B",
    volume = "631",
    pages = "48--57",
    year = "2005"
}

@article{Bonvini:2014joa,
    author = "Bonvini, Marco and Marzani, Simone",
    title = "{Resummed Higgs cross section at N$^{3}$LL}",
    eprint = "1405.3654",
    archivePrefix = "arXiv",
    primaryClass = "hep-ph",
    reportNumber = "DESY-14-075, DCPT-14-94, IPPP-14-47",
    doi = "10.1007/JHEP09(2014)007",
    journal = "JHEP",
    volume = "09",
    pages = "007",
    year = "2014"
}

@article{Bonvini:2016frm,
    author = "Bonvini, Marco and Marzani, Simone and Muselli, Claudio and Rottoli, Luca",
    title = "{On the Higgs cross section at N$^{3}$LO+N$^{3}$LL and its uncertainty}",
    eprint = "1603.08000",
    archivePrefix = "arXiv",
    primaryClass = "hep-ph",
    reportNumber = "OUTP-16-05P, TIF-UNIMI-2016-2",
    doi = "10.1007/JHEP08(2016)105",
    journal = "JHEP",
    volume = "08",
    pages = "105",
    year = "2016"
}

@article{Idilbi:2006dg,
    author = "Idilbi, Ahmad and Ji, Xiang-dong and Yuan, Feng",
    title = "{Resummation of threshold logarithms in effective field theory for DIS, Drell-Yan and Higgs production}",
    eprint = "hep-ph/0605068",
    archivePrefix = "arXiv",
    doi = "10.1016/j.nuclphysb.2006.07.002",
    journal = "Nucl. Phys. B",
    volume = "753",
    pages = "42--68",
    year = "2006"
}

@article{Moch:2005ba,
    author = "Moch, S. and Vermaseren, J. A. M. and Vogt, A.",
    title = "{Higher-order corrections in threshold resummation}",
    eprint = "hep-ph/0506288",
    archivePrefix = "arXiv",
    reportNumber = "DESY-05-105, SFB-CPP-05-25, DCPT-05-60, IPPP-05-30, NIKHEF-05-010",
    doi = "10.1016/j.nuclphysb.2005.08.005",
    journal = "Nucl. Phys. B",
    volume = "726",
    pages = "317--335",
    year = "2005"
}

@article{Becher:2006mr,
    author = "Becher, Thomas and Neubert, Matthias and Pecjak, Ben D.",
    title = "{Factorization and Momentum-Space Resummation in Deep-Inelastic Scattering}",
    eprint = "hep-ph/0607228",
    archivePrefix = "arXiv",
    reportNumber = "CLNS-06-1971, FERMILAB-PUB-06-242-T, DFB-CPP-06-32, SI-HEP-2006-09",
    doi = "10.1088/1126-6708/2007/01/076",
    journal = "JHEP",
    volume = "01",
    pages = "076",
    year = "2007"
}

@article{Ravindran:2006cg,
    author = "Ravindran, V.",
    title = "{Higher-order threshold effects to inclusive processes in QCD}",
    eprint = "hep-ph/0603041",
    archivePrefix = "arXiv",
    reportNumber = "HRI-03-2006",
    doi = "10.1016/j.nuclphysb.2006.06.025",
    journal = "Nucl. Phys. B",
    volume = "752",
    pages = "173--196",
    year = "2006"
}

@article{Ravindran:2005vv,
    author = "Ravindran, V.",
    title = "{On Sudakov and soft resummations in QCD}",
    eprint = "hep-ph/0512249",
    archivePrefix = "arXiv",
    reportNumber = "HRI-12-2005",
    doi = "10.1016/j.nuclphysb.2006.04.008",
    journal = "Nucl. Phys. B",
    volume = "746",
    pages = "58--76",
    year = "2006"
}

@article{Catani:1996yz,
    author = "Catani, Stefano and Mangano, Michelangelo L. and Nason, Paolo and Trentadue, Luca",
    title = "{The Resummation of soft gluons in hadronic collisions}",
    eprint = "hep-ph/9604351",
    archivePrefix = "arXiv",
    reportNumber = "CERN-TH-96-86",
    doi = "10.1016/0550-3213(96)00399-9",
    journal = "Nucl. Phys. B",
    volume = "478",
    pages = "273--310",
    year = "1996"
}

@article{Harlander:2014wda,
    author = "Harlander, Robert V. and Kulesza, Anna and Theeuwes, Vincent and Zirke, Tom",
    title = "{Soft gluon resummation for gluon-induced Higgs Strahlung}",
    eprint = "1410.0217",
    archivePrefix = "arXiv",
    primaryClass = "hep-ph",
    reportNumber = "MS-TP-14-17, WUB-14-10, LPN14-112",
    doi = "10.1007/JHEP11(2014)082",
    journal = "JHEP",
    volume = "11",
    pages = "082",
    year = "2014"
}

@article{Dawson:2012gs,
    author = "Dawson, S. and Han, T. and Lai, W. K. and Leibovich, A. K. and Lewis, I.",
    title = "{Resummation Effects in Vector-Boson and Higgs Associated Production}",
    eprint = "1207.4207",
    archivePrefix = "arXiv",
    primaryClass = "hep-ph",
    doi = "10.1103/PhysRevD.86.074007",
    journal = "Phys. Rev. D",
    volume = "86",
    pages = "074007",
    year = "2012"
}

@article{Ahmed:2020nci,
    author = "Ahmed, Taushif and Ajjath, A. H. and Das, Goutam and Mukherjee, Pooja and Ravindran, V. and Tiwari, Surabhi",
    title = "{Soft-virtual correction and threshold resummation for $n$-colorless particles to fourth order in QCD: Part I}",
    eprint = "2010.02979",
    archivePrefix = "arXiv",
    primaryClass = "hep-ph",
    reportNumber = "MPP-2020-51, SI-HEP-2020-02",
    month = "10",
    year = "2020"
}

@article{Ahmed:2016otz,
    author = "Ahmed, Taushif and Bonvini, Marco and Kumar, M. C. and Mathews, Prakash and Rana, Narayan and Ravindran, V. and Rottoli, Luca",
    title = "{Pseudo-scalar Higgs boson production at N$^3$ LO$_{\text {A}}$ +N$^3$ LL $'$}",
    eprint = "1606.00837",
    archivePrefix = "arXiv",
    primaryClass = "hep-ph",
    reportNumber = "OUTP-16-13P",
    doi = "10.1140/epjc/s10052-016-4510-1",
    journal = "Eur. Phys. J. C",
    volume = "76",
    number = "12",
    pages = "663",
    year = "2016"
}

@article{Schmidt:2015cea,
    author = "Schmidt, Timo and Spira, Michael",
    title = "{Higgs Boson Production via Gluon Fusion: Soft-Gluon Resummation including Mass Effects}",
    eprint = "1509.00195",
    archivePrefix = "arXiv",
    primaryClass = "hep-ph",
    reportNumber = "PSI-PR-15-08",
    doi = "10.1103/PhysRevD.93.014022",
    journal = "Phys. Rev. D",
    volume = "93",
    number = "1",
    pages = "014022",
    year = "2016"
}

@article{deFlorian:2007sr,
    author = "de Florian, Daniel and Zurita, Jose",
    title = "{Soft-gluon resummation for pseudoscalar Higgs boson production at hadron colliders}",
    eprint = "0711.1916",
    archivePrefix = "arXiv",
    primaryClass = "hep-ph",
    doi = "10.1016/j.physletb.2007.11.018",
    journal = "Phys. Lett. B",
    volume = "659",
    pages = "813--820",
    year = "2008"
}

@article{Das:2019bxi,
    author = "Das, Goutam and Kumar, M. C. and Samanta, Kajal",
    title = "{Resummed inclusive cross-section in ADD model at N$^{3}$LL}",
    eprint = "1912.13039",
    archivePrefix = "arXiv",
    primaryClass = "hep-ph",
    reportNumber = "DESY-19-171, DESY 19-171",
    doi = "10.1007/JHEP10(2020)161",
    journal = "JHEP",
    volume = "10",
    pages = "161",
    year = "2020"
}

@article{Das:2020gie,
    author = "Das, Goutam and Kumar, M. C. and Samanta, Kajal",
    title = "{Resummed inclusive cross-section in Randall-Sundrum model at NNLO+NNLL}",
    eprint = "2004.03938",
    archivePrefix = "arXiv",
    primaryClass = "hep-ph",
    doi = "10.1007/JHEP07(2020)040",
    journal = "JHEP",
    volume = "07",
    pages = "040",
    year = "2020"
}

@article{Das:2020pzo,
    author = "Das, Goutam and Kumar, M. C. and Samanta, Kajal",
    title = "{Precision QCD phenomenology of exotic spin-2 search at the LHC}",
    eprint = "2011.15121",
    archivePrefix = "arXiv",
    primaryClass = "hep-ph",
    reportNumber = "SI-HEP-2020-32, P3H-20-077",
    doi = "10.1007/JHEP04(2021)111",
    journal = "JHEP",
    volume = "04",
    pages = "111",
    year = "2021"
}

@article{Vogt:2004ns,
    author = "Vogt, A.",
    title = "{Efficient evolution of unpolarized and polarized parton distributions with QCD-PEGASUS}",
    eprint = "hep-ph/0408244",
    archivePrefix = "arXiv",
    reportNumber = "NIKHEF-04-011",
    doi = "10.1016/j.cpc.2005.03.103",
    journal = "Comput. Phys. Commun.",
    volume = "170",
    pages = "65--92",
    year = "2005"
}

@article{Buckley:2014ana,
    author = {Buckley, Andy and Ferrando, James and Lloyd, Stephen and Nordstr\"om, Karl and Page, Ben and R\"ufenacht, Martin and Sch\"onherr, Marek and Watt, Graeme},
    title = "{LHAPDF6: parton density access in the LHC precision era}",
    eprint = "1412.7420",
    archivePrefix = "arXiv",
    primaryClass = "hep-ph",
    reportNumber = "GLAS-PPE-2014-05, MCNET-14-29, IPPP-14-111, DCPT-14-222",
    doi = "10.1140/epjc/s10052-015-3318-8",
    journal = "Eur. Phys. J. C",
    volume = "75",
    pages = "132",
    year = "2015"
}

@article{Brein:2012ne,
    author = "Brein, Oliver and Harlander, Robert V. and Zirke, Tom J. E.",
    title = "{vh@nnlo - Higgs Strahlung at hadron colliders}",
    eprint = "1210.5347",
    archivePrefix = "arXiv",
    primaryClass = "hep-ph",
    doi = "10.1016/j.cpc.2012.11.002",
    journal = "Comput. Phys. Commun.",
    volume = "184",
    pages = "998--1003",
    year = "2013"
}

@article{Campbell:2016jau,
    author = "Campbell, John M. and Ellis, R. Keith and Williams, Ciaran",
    title = "{Associated production of a Higgs boson at NNLO}",
    eprint = "1601.00658",
    archivePrefix = "arXiv",
    primaryClass = "hep-ph",
    reportNumber = "IPPP-15-78, FERMILAB-PUB-16-001-T",
    doi = "10.1007/JHEP06(2016)179",
    journal = "JHEP",
    volume = "06",
    pages = "179",
    year = "2016"
}

@article{Brein:2011vx,
    author = "Brein, Oliver and Harlander, Robert and Wiesemann, Marius and Zirke, Tom",
    title = "{Top-Quark Mediated Effects in Hadronic Higgs-Strahlung}",
    eprint = "1111.0761",
    archivePrefix = "arXiv",
    primaryClass = "hep-ph",
    reportNumber = "CERN-PH-TH-2011-268, FR-PHENO-2011-016, WUB-11-15",
    doi = "10.1140/epjc/s10052-012-1868-6",
    journal = "Eur. Phys. J. C",
    volume = "72",
    pages = "1868",
    year = "2012"
}

@article{Mueller:1979ih,
    author = "Mueller, Alfred H.",
    title = "{On the Asymptotic Behavior of the Sudakov Form-factor}",
    reportNumber = "CU-TP-158",
    doi = "10.1103/PhysRevD.20.2037",
    journal = "Phys. Rev. D",
    volume = "20",
    pages = "2037",
    year = "1979"
}

@article{Sudakov:1954sw,
    author = "Sudakov, V. V.",
    title = "{Vertex parts at very high-energies in quantum electrodynamics}",
    journal = "Sov. Phys. JETP",
    volume = "3",
    pages = "65--71",
    year = "1956"
}

@article{Collins:1980ih,
    author = "Collins, John C.",
    title = "{Algorithm to Compute Corrections to the Sudakov Form-factor}",
    reportNumber = "Print-80-0421 (PRINCETON)",
    doi = "10.1103/PhysRevD.22.1478",
    journal = "Phys. Rev. D",
    volume = "22",
    pages = "1478",
    year = "1980"
}

@article{Sen:1981sd,
    author = "Sen, Ashoke",
    title = "{Asymptotic Behavior of the Sudakov Form-Factor in QCD}",
    reportNumber = "ITP-SB-81-19",
    doi = "10.1103/PhysRevD.24.3281",
    journal = "Phys. Rev. D",
    volume = "24",
    pages = "3281",
    year = "1981"
}

@article{Davies:2020drs,
    author = "Davies, Joshua and Mishima, Go and Steinhauser, Matthias",
    title = "{Virtual corrections to $gg\to ZH$ in the high-energy and large-$m_t$ limits}",
    eprint = "2011.12314",
    archivePrefix = "arXiv",
    primaryClass = "hep-ph",
    reportNumber = "TTP20-041, P3H-20-074",
    doi = "10.1007/JHEP03(2021)034",
    journal = "JHEP",
    volume = "03",
    pages = "034",
    year = "2021"
}

@article{Laenen:2005uz,
    author = "Laenen, Eric and Magnea, Lorenzo",
    title = "{Threshold resummation for electroweak annihilation from DIS data}",
    eprint = "hep-ph/0508284",
    archivePrefix = "arXiv",
    reportNumber = "DFTT-26-2005, NIKHEF-2005-015",
    doi = "10.1016/j.physletb.2005.10.038",
    journal = "Phys. Lett. B",
    volume = "632",
    pages = "270--276",
    year = "2006"
}

@article{Moch:2004pa,
    author = "Moch, S. and Vermaseren, J. A. M. and Vogt, A.",
    title = "{The Three loop splitting functions in QCD: The Nonsinglet case}",
    eprint = "hep-ph/0403192",
    archivePrefix = "arXiv",
    reportNumber = "DESY-04-047, SFB-CPP-04-09, NIKHEF-04-001",
    doi = "10.1016/j.nuclphysb.2004.03.030",
    journal = "Nucl. Phys. B",
    volume = "688",
    pages = "101--134",
    year = "2004"
}

@article{Vogt:2004mw,
    author = "Vogt, A. and Moch, S. and Vermaseren, J. A. M.",
    title = "{The Three-loop splitting functions in QCD: The Singlet case}",
    eprint = "hep-ph/0404111",
    archivePrefix = "arXiv",
    reportNumber = "NIKHEF-04-004, DESY-04-060, SFB-CPP-04-12",
    doi = "10.1016/j.nuclphysb.2004.04.024",
    journal = "Nucl. Phys. B",
    volume = "691",
    pages = "129--181",
    year = "2004"
}

@article{ATLAS:2024yzu,
    author = "Aad, Georges and others",
    collaboration = "ATLAS",
    title = "{Measurements of $WH$ and $ZH$ production with Higgs boson decays into bottom quarks and direct constraints on the charm Yukawa coupling in $13\,\mathrm{TeV}$$pp$ collisions with the ATLAS detector}",
    eprint = "2410.19611",
    archivePrefix = "arXiv",
    primaryClass = "hep-ex",
    reportNumber = "CERN-EP-2024-237",
    month = "10",
    year = "2024"
}

@article{Kniehl:2011aa,
    author = "Kniehl, Bernd A. and Palisoc, Caesar P.",
    title = "{Associated production of Z and neutral Higgs bosons at the CERN Large Hadron Collider}",
    eprint = "1112.1575",
    archivePrefix = "arXiv",
    primaryClass = "hep-ph",
    reportNumber = "DESY-11-239",
    doi = "10.1103/PhysRevD.85.075027",
    journal = "Phys. Rev. D",
    volume = "85",
    pages = "075027",
    year = "2012"
}

@article{Goncalves:2015mfa,
    author = {Goncalves, Dorival and Krauss, Frank and Kuttimalai, Silvan and Maierh\"ofer, Philipp},
    title = "{Higgs-Strahlung: Merging the NLO Drell-Yan and Loop-Induced 0+1 jet Multiplicities}",
    eprint = "1509.01597",
    archivePrefix = "arXiv",
    primaryClass = "hep-ph",
    reportNumber = "IPPP-15-57, DCPT-15-114",
    doi = "10.1103/PhysRevD.92.073006",
    journal = "Phys. Rev. D",
    volume = "92",
    number = "7",
    pages = "073006",
    year = "2015"
}

@article{Catani:1990rp,
    author = "Catani, S. and Trentadue, L.",
    title = "{Comment on QCD exponentiation at large x}",
    reportNumber = "CAVENDISH-HEP-90-12",
    doi = "10.1016/0550-3213(91)90506-S",
    journal = "Nucl. Phys. B",
    volume = "353",
    pages = "183--186",
    year = "1991"
}

@article{Ahmed:2015sna,
    author = "Ahmed, Taushif and Das, Goutam and Kumar, M. C. and Rana, Narayan and Ravindran, V.",
    title = "{RG improved Higgs boson production to N$^3$LO in QCD}",
    eprint = "1505.07422",
    archivePrefix = "arXiv",
    primaryClass = "hep-ph",
    month = "5",
    year = "2015"
}

@article{AH:2019phz,
    author = "A H, Ajjath and Chakraborty, Amlan and Das, Goutam and Mukherjee, Pooja and Ravindran, V.",
    title = "{Resummed prediction for Higgs boson production through b$ \overline{\mathrm{b}} $ annihilation at N$^{3}$LL}",
    eprint = "1905.03771",
    archivePrefix = "arXiv",
    primaryClass = "hep-ph",
    reportNumber = "IMSC/2019/05/04, DESY-19-076, DESY 19-076",
    doi = "10.1007/JHEP11(2019)006",
    journal = "JHEP",
    volume = "11",
    pages = "006",
    year = "2019"
}

@article{AH:2020cok,
    author = "A H, Ajjath and Das, Goutam and Kumar, M. C. and Mukherjee, Pooja and Ravindran, V. and Samanta, Kajal",
    title = "{Resummed Drell-Yan cross-section at N$^{3}$LL}",
    eprint = "2001.11377",
    archivePrefix = "arXiv",
    primaryClass = "hep-ph",
    reportNumber = "IMSc/2019/12/16, DESY-19-191, SI-HEP-2019-23",
    doi = "10.1007/JHEP10(2020)153",
    journal = "JHEP",
    volume = "10",
    pages = "153",
    year = "2020"
}

@article{AH:2020iki,
    author = "A H, Ajjath and Mukherjee, Pooja and Ravindran, V.",
    title = "{Next to soft corrections to Drell-Yan and Higgs boson productions}",
    eprint = "2006.06726",
    archivePrefix = "arXiv",
    primaryClass = "hep-ph",
    doi = "10.1103/PhysRevD.105.094035",
    journal = "Phys. Rev. D",
    volume = "105",
    number = "9",
    pages = "094035",
    year = "2022"
}

@article{AH:2021vdc,
    author = "A H, Ajjath and Mukherjee, Pooja and Ravindran, V. and Sankar, Aparna and Tiwari, Surabhi",
    title = "{Resummed Higgs boson cross section at next-to SV to ${\mathrm{NNLO}}+ {\overline{\mathrm{NNLL}}}$}",
    eprint = "2109.12657",
    archivePrefix = "arXiv",
    primaryClass = "hep-ph",
    reportNumber = "IMSc/2021/09/06",
    doi = "10.1140/epjc/s10052-022-10752-9",
    journal = "Eur. Phys. J. C",
    volume = "82",
    number = "9",
    pages = "774",
    year = "2022"
}

@article{vanBeekveld:2021hhv,
    author = "van Beekveld, Melissa and Laenen, Eric and Sinninghe Damst\'e, Jort and Vernazza, Leonardo",
    title = "{Next-to-leading power threshold corrections for finite order and resummed colour-singlet cross sections}",
    eprint = "2101.07270",
    archivePrefix = "arXiv",
    primaryClass = "hep-ph",
    reportNumber = "Nikhef/2021-001",
    doi = "10.1007/JHEP05(2021)114",
    journal = "JHEP",
    volume = "05",
    pages = "114",
    year = "2021"
}

@article{Banerjee:2024xdh,
    author = "Banerjee, Pulak and Dey, Chinmoy and Kumar, M. C. and Pandey, Vaibhav",
    title = "{Threshold resummation for $Z$-boson pair production at NNLO+NNLL}",
    eprint = "2409.16375",
    archivePrefix = "arXiv",
    primaryClass = "hep-ph",
    month = "9",
    year = "2024"
}

@article{PDF4LHCWorkingGroup:2022cjn,
    author = "Ball, Richard D. and others",
    collaboration = "PDF4LHC Working Group",
    title = "{The PDF4LHC21 combination of global PDF fits for the LHC Run III}",
    eprint = "2203.05506",
    archivePrefix = "arXiv",
    primaryClass = "hep-ph",
    reportNumber = "Edinburgh 2021/31, FERMILAB-PUB-22-121-QIS-SCD-T, MSUHEP-22-010,
  Nikhef 2021-033, SMU-HEP-22-01",
    doi = "10.1088/1361-6471/ac7216",
    journal = "J. Phys. G",
    volume = "49",
    number = "8",
    pages = "080501",
    year = "2022"
}

@article{ParticleDataGroup:2024cfk,
    author = "Navas, S. and others",
    collaboration = "Particle Data Group",
    title = "{Review of particle physics}",
    doi = "10.1103/PhysRevD.110.030001",
    journal = "Phys. Rev. D",
    volume = "110",
    number = "3",
    pages = "030001",
    year = "2024"
}

@article{Moch:2009hr,
    author = "Moch, S. and Vogt, A.",
    title = "{On non-singlet physical evolution kernels and large-x coefficient functions in perturbative QCD}",
    eprint = "0909.2124",
    archivePrefix = "arXiv",
    primaryClass = "hep-ph",
    reportNumber = "DESY-09-133, SFB-CPP-09-80, LTH-840",
    doi = "10.1088/1126-6708/2009/11/099",
    journal = "JHEP",
    volume = "11",
    pages = "099",
    year = "2009"
}

@article{Soar:2009yh,
    author = "Soar, G. and Moch, S. and Vermaseren, J. A. M. and Vogt, A.",
    title = "{On Higgs-exchange DIS, physical evolution kernels and fourth-order splitting functions at large x}",
    eprint = "0912.0369",
    archivePrefix = "arXiv",
    primaryClass = "hep-ph",
    reportNumber = "LTH-857, DESY-09-211, SFB-CPP-09-119, NIKHEF-09-031",
    doi = "10.1016/j.nuclphysb.2010.02.003",
    journal = "Nucl. Phys. B",
    volume = "832",
    pages = "152--227",
    year = "2010"
}

@article{Bonocore:2015esa,
    author = "Bonocore, D. and Laenen, E. and Magnea, L. and Melville, S. and Vernazza, L. and White, C. D.",
    title = "{A factorization approach to next-to-leading-power threshold logarithms}",
    eprint = "1503.05156",
    archivePrefix = "arXiv",
    primaryClass = "hep-ph",
    reportNumber = "EDINBURGH-2015-03, NIKHEF-2015-005, ITF-UU-15-02",
    doi = "10.1007/JHEP06(2015)008",
    journal = "JHEP",
    volume = "06",
    pages = "008",
    year = "2015"
}

@article{DelDuca:2017twk,
    author = "Del Duca, V. and Laenen, E. and Magnea, L. and Vernazza, L. and White, C. D.",
    title = "{Universality of next-to-leading power threshold effects for colourless final states in hadronic collisions}",
    eprint = "1706.04018",
    archivePrefix = "arXiv",
    primaryClass = "hep-ph",
    reportNumber = "NIKHEF-2017-25, EDINBURGH-2017-10, QMUL-PH-17-07, ARC-17-03",
    doi = "10.1007/JHEP11(2017)057",
    journal = "JHEP",
    volume = "11",
    pages = "057",
    year = "2017"
}

@article{Bahjat-Abbas:2019fqa,
    author = "Bahjat-Abbas, N. and Bonocore, D. and Sinninghe Damst\'e, J. and Laenen, E. and Magnea, L. and Vernazza, L. and White, C. D.",
    title = "{Diagrammatic resummation of leading-logarithmic threshold effects at next-to-leading power}",
    eprint = "1905.13710",
    archivePrefix = "arXiv",
    primaryClass = "hep-ph",
    reportNumber = "QMUL-PH-19-12, Nikhef/2019-015, MS-TP-19-08",
    doi = "10.1007/JHEP11(2019)002",
    journal = "JHEP",
    volume = "11",
    pages = "002",
    year = "2019"
}

@article{Beneke:2019oqx,
    author = "Beneke, Martin and Broggio, Alessandro and Jaskiewicz, Sebastian and Vernazza, Leonardo",
    title = "{Threshold factorization of the Drell-Yan process at next-to-leading power}",
    eprint = "1912.01585",
    archivePrefix = "arXiv",
    primaryClass = "hep-ph",
    reportNumber = "TUM-HEP-1239/19",
    doi = "10.1007/JHEP07(2020)078",
    journal = "JHEP",
    volume = "07",
    pages = "078",
    year = "2020"
}

@article{vanBeekveld:2019cks,
    author = "van Beekveld, Melissa and Beenakker, Wim and Basu, Rahul and Laenen, Eric and Misra, Anuradha and Motylinski, Patrick",
    title = "{Next-to-leading power threshold effects for resummed prompt photon production}",
    eprint = "1905.11771",
    archivePrefix = "arXiv",
    primaryClass = "hep-ph",
    reportNumber = "FR-PHENO-2019-008; Nikhef/2019-010",
    doi = "10.1103/PhysRevD.100.056009",
    journal = "Phys. Rev. D",
    volume = "100",
    number = "5",
    pages = "056009",
    year = "2019"
}

@article{AH:2022lpp,
    author = "A H, Ajjath and Mukherjee, Pooja and Ravindran, V.",
    title = "{Going beyond soft plus virtual}",
    eprint = "2204.09012",
    archivePrefix = "arXiv",
    primaryClass = "hep-ph",
    doi = "10.1103/PhysRevD.105.L091503",
    journal = "Phys. Rev. D",
    volume = "105",
    number = "9",
    pages = "L091503",
    year = "2022"
}

@article{Granata:2017iod,
    author = "Granata, Federico and Lindert, Jonas M. and Oleari, Carlo and Pozzorini, Stefano",
    title = "{NLO QCD+EW predictions for HV and HV +jet production including parton-shower effects}",
    eprint = "1706.03522",
    archivePrefix = "arXiv",
    primaryClass = "hep-ph",
    doi = "10.1007/JHEP09(2017)012",
    journal = "JHEP",
    volume = "09",
    pages = "012",
    year = "2017"
}


\end{document}